\documentclass[reprint, superscriptaddress, secnumarabic, amssymb, nobibnotes, aps, prl]{revtex4-1}

\usepackage{chemformula,siunitx}
\usepackage{lastpage}

\usepackage[protrusion=true, expansion=true]{microtype}
\usepackage{wrapfig}
\usepackage{multirow}
\usepackage{array}
\usepackage{ragged2e}
\justifying

\usepackage{epstopdf}
\usepackage[T1]{fontenc}
\usepackage{amsbsy}
\usepackage[T1]{fontenc}
\usepackage{amsmath}
\usepackage{amssymb}
\usepackage{bbm}
\usepackage{braket}
\usepackage{xcolor}
\allowdisplaybreaks
\usepackage{graphicx}

\usepackage[normalem]{ulem}

\renewcommand{\approx}{\simeq}

\usepackage[colorlinks=true]{hyperref}
\hypersetup{
    unicode=false,          % non-Latin characters 
    pdftoolbar=true,        % show Acrobat
    pdfmenubar=true,        % show Acrobat 
    pdffitwindow=false,     % window fit to page when opened
    pdfstartview={FitH},    % fits the width of the page to the window
    pdftitle={SnAs},    % title
    pdfauthor={Shashank Srivastava},      %author
    pdfsubject={},   % subject of the document
    pdfcreator={},   % creator of the document
    pdfproducer={}, % producer of the document
    pdfkeywords={} {} {}, % list of keywords
    pdfnewwindow=true,      % links in new window
    colorlinks=true,       % false: boxed links; true: colored links
    linkcolor=blue, %red,          % color of internal links (change box color with linkbordercolor)
    citecolor=blue,        % color of links to bibliography
    filecolor=magenta,      % color of file links
    urlcolor=blue          % color of external links
}

\begin{document}

\title{Probing the intermediate state of type-I superconductor SnAs using Muon Spin Spectroscopy}

\author{Shashank Srivastava}
\affiliation{Department of Physics, Indian Institute of Science Education and Research Bhopal, Bhopal, 462066, India}
\author{Omkar Kulkarni}
\affiliation{Department of Physics, Indian Institute of Science Education and Research Bhopal, Bhopal, 462066, India}
\author{Arushi}
\affiliation{Department of Physics, Indian Institute of Science Education and Research Bhopal, Bhopal, 462066, India}
\author{Deepak Singh}
\affiliation{ISIS Facility, STFC Rutherford Appleton Laboratory, Oxfordshire, OX11 0QX, United Kingdom}
\author{Poulami Manna}
\affiliation{Department of Physics, Indian Institute of Science Education and Research Bhopal, Bhopal, 462066, India}
\author{Priya Mishra}
\affiliation{Department of Physics, Indian Institute of Science Education and Research Bhopal, Bhopal, 462066, India}
\author{Suhani Sharma}
\affiliation{Department of Physics, Indian Institute of Science Education and Research Bhopal, Bhopal, 462066, India}
\author{Pabitra Kumar Biswas}\thanks{Deceased.}
\affiliation{ISIS Facility, STFC Rutherford Appleton Laboratory, Oxfordshire, OX11 0QX, United Kingdom}
\author{Rhea Stewart}
\affiliation{ISIS Facility, STFC Rutherford Appleton Laboratory, Oxfordshire, OX11 0QX, United Kingdom}
\author{Adrian D. Hillier}
\affiliation{ISIS Facility, STFC Rutherford Appleton Laboratory, Oxfordshire, OX11 0QX, United Kingdom}
\author{Ravi Prakash Singh}
\email[]{rpsingh@iiserb.ac.in}
\affiliation{Department of Physics, Indian Institute of Science Education and Research Bhopal, Bhopal, 462066, India}

\begin{abstract}
Superconductivity with non-trivial band topology provides a novel platform for exploring topological superconductivity and its quantum applications. A detailed microscopic understanding of the superconducting ground state in such materials is crucial. Here, we report the results of a muon spin rotation/relaxation study ($\mu$SR) of the topologically non-trivial superconductor SnAs, which exhibits superconductivity below 3.74(1) \si{K}. Zero-field (ZF) $\mu$SR data reveal that this system is a time-reversal invariant superconductor, and systematic transverse-field (TF) $\mu$SR measurements unveil the type-I nature of the SnAs superconductor. We have established the superconducting phase diagram to understand the intermediate state of type-I superconductors. Moreover, ab \textit{initio} band structure and phonon calculations are performed, which correlate with the experimental characterization.
\end{abstract}
\maketitle

\section{INTRODUCTION}
Topological systems are known for their non-trivial band structures, which have a significant impact on the electronic properties of the material. Non-trivial behavior leads to the formation of robust surface states that are immune to small perturbations such as disorder \cite{topomat,viewpoint}. A well-known topological system that contains gapless electronic phases is the topological semimetal, notable for its energy bands with topologically stable intersections \cite{burkov2016topological,armitage2018weyl,gao2019tsm}. The search for superconductivity in intrinsic topological materials provides an exciting avenue to discover topological superconductors \cite{wang2018tsm,Sato,nayak2008topo,qi2011tis}. These materials promise transformative advances in fundamental physics and quantum technology applications \cite{alicea2011non,liu2014prx}. One of the vital factors for the emergence of a topological superconducting state is strong spin-orbit coupling (SOC), which can lead to the emergence of odd-parity pairing, enabling the realization of time-reversal invariant topological superconductors \cite{Sato,TSCinTM,cuxbi2se3}. 

The binary series of compounds, SnX (X$=$P, As, Sb, S, Se, Te) has attracted much attention due to their non-trivial band topology \cite{SnP,sntetopological,SnSeSnS,snsbx,snte}. Among them, the compounds SnSb and SnAs have inherent superconducting properties and have been candidates for topological superconductivity \cite{naclprl,sn0.4sb0.6,snasprb,snastypeII}. Recent theoretical studies suggest that SnAs is a topological semimetal, characterized by considerable band splitting \cite{snastypeII,snasprb,snasarpes}. SnAs share similar characteristics of type-I superconductors with strong SOC, making them a promising candidate for topological superconductivity \cite{irsn4dft,pdte2,pdte2type1,irsn4}. The origin of superconductivity in SnAs—whether it arises from a mixed-valence state of Sn (Sn$^{4+}$ and Sn$^{2+}$) or a single valence configuration of Sn$^{3+}$ —remains an open question \cite{mixedvalence,snasvalence}. Although several studies report a conventional s-wave superconducting gap symmetry in SnAs, its classification as a type-I or type-II superconductor is still under debate, highlighting the need for further microscopic investigations \cite{snasprb,snasvalence,snastypeII,snasheatcap,snassheet}. 

Muon studies on type-I superconductors (\cite{al6re,lipd2ge,irsn4,aube,Sntype1,Pb2Pd,pdte2type1}, and refs. therein) are limited, and only some reports have probed the intermediate state with their superconducting phase diagrams \cite{Sntype1,irsn4,aube}. At the same time, superconducting ground state studies of these materials having non-trivial band topology are still missing, despite their importance for understanding topological superconductivity. 

To address this gap and clarify the superconducting pairing mechanism, we performed zero-field (ZF) and transverse-field (TF) muon spin rotation/relaxation ($\mu$SR) measurements on high-quality single crystals of SnAs. This technique is a powerful probe that directly enables the study of various superconducting regions— Meissner, intermediate, mixed and normal — and can easily explore the coexistence of type-I and type-II superconductivity\cite{aube,Pb2Pd,pdte2type1,Sntype1,zrb12coexistence,nbge2coexistence,pdte2coexistence,LaCuSb2}. Our results unambiguously establish conventional type-I superconductivity in SnAs, successfully resolving the existing debate regarding a possible mixture of type-I and type-II behavior. The preservation of time-reversal symmetry and the observed superconducting response further support a single valence state of Sn, as a mixed valence would be expected to yield an unconventional superconducting ground state \cite{AgSnSe2,valencetaraphder,valencevarma}. In addition, we construct the complete superconducting phase diagram, and perform complementary band-structure and phonon calculations, which further substantiate the presence of a non-trivial band topology in SnAs.

\begin{figure*}
\includegraphics[width=2.05\columnwidth]{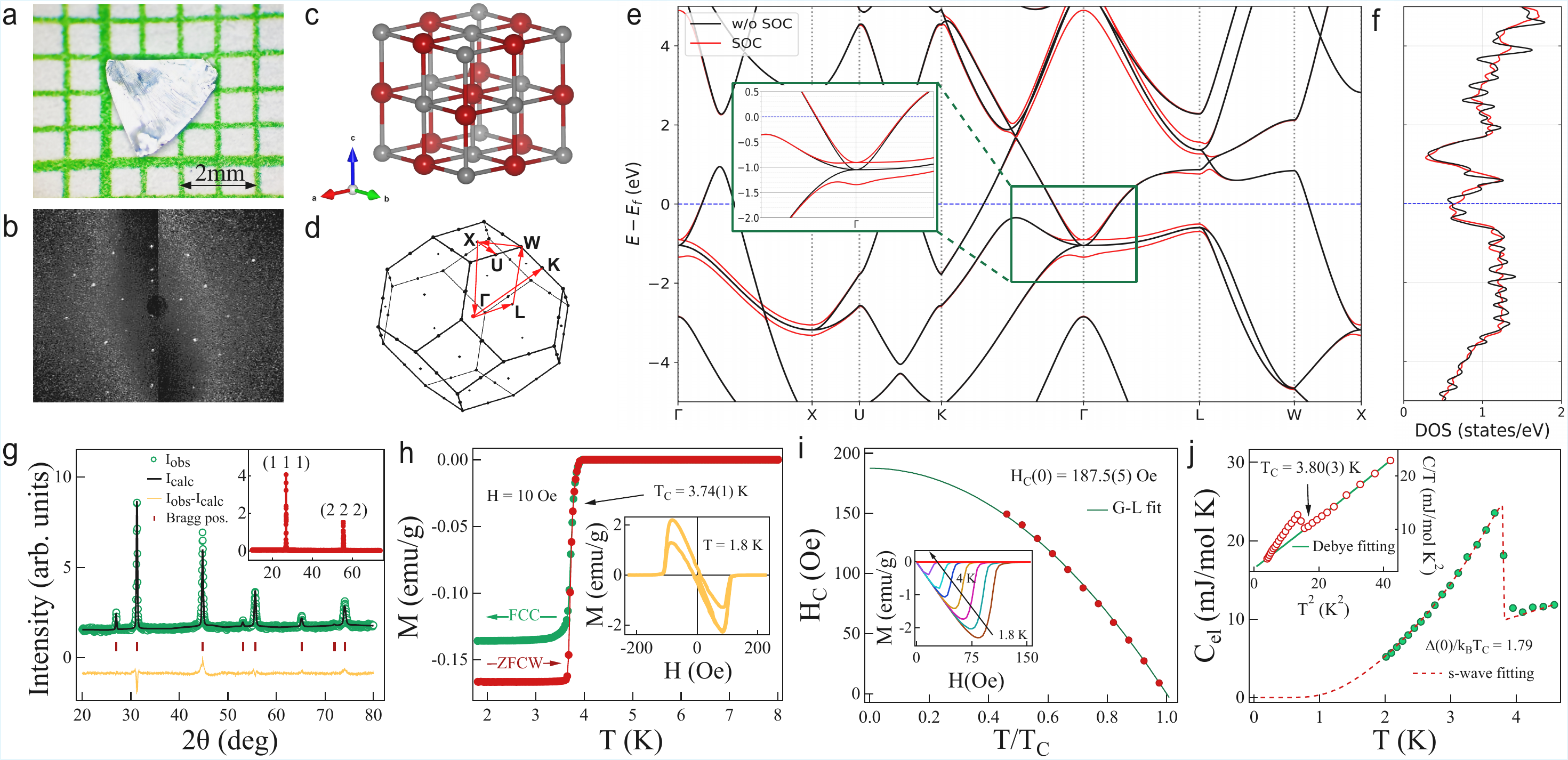}
\caption {\label{Fig1} (a) Microscopic image and (b) Laue diffraction pattern for SnAs single crystal. (c) Crystal structure of the SnAs unit cell. Grey and maroon balls denote the As and the Sn atoms, respectively. (d) The first Brillouin zone with relevant high-symmetry directions is shown using red arrows. (e,f) The bulk band structure and electronic DOS, with and without the SOC contribution. Inset of (e) shows the zoomed view of the band degeneracy at the $\Gamma$ point. (g) Rietveld refinement of the powder XRD spectrum. Inset: XRD pattern for the SnAs single crystal. (h) Temperature-dependent magnetization indicates $T_C$ at 3.74(1) \si{K}. The inset shows the M-H loop at 1.8 \si{K}. (i) Critical field versus the normalized temperature is fitted with the GL equation, giving an $H_C$ of 187.5(5) \si{Oe}. Inset: Magnetization data at low applied field at various temperatures below $T_C$. (j) Electronic specific heat fitted with the s-wave model. Inset displays the Debye fitting for the normal state $C/T$ versus $T^2$ data.}
% The five-quadrant M versus H loop is shown in the inset. 
\end{figure*}

\section{RESULTS AND DISCUSSION}
\subsection{Sample Characterization}
The powder XRD pattern for the SnAs crystals was Rietveld refined using the FullProf Suite software \cite{fullprof} to ensure phase purity (Fig. \ref{Fig1}(g)). It crystallized into the rock salt crystal structure (space group $Fm\bar3m$) with $a=5.7221(5)~\text{\AA}$ and $V_{cell}=187.35(5)~\text{\AA}^3$. The face-centered cubic structure of the SnAs was constructed with the VESTA software \cite{momma2011vesta} as shown in Fig. \ref{Fig1}(c). Energy Dispersive X-ray (EDX) analysis confirms the nominal composition of the compound (Supplemental Material (SM) \cite{SM}, Fig. \textcolor{blue}{S1}(a)). The Laue diffraction pattern (Fig. \ref{Fig1}(b)) and the Selected Area Electron Diffraction (SAED) pattern (SM \cite{SM}, Fig. \textcolor{blue}{S1}(b)) confirm the high-quality single-crystalline nature of the sample. The crystallinity of the material was confirmed by XRD on the single crystals as depicted in the inset of Fig. \ref{Fig1}(g).

\subsection{Magnetization and Specific Heat}
The DC magnetization and specific heat measurements demonstrate the bulk superconductivity in SnAs. The diamagnetic transition in magnetization at 3.74(1) \si{K} is shown in Fig. \ref{Fig1}(h). The magnetic field-dependent magnetization data at different temperatures exhibit a sharp transition, similar to that of type-I superconductors. Data can be fitted with the Ginzburg-Landau (G-L) equation (see details in SM \cite{SM}), to obtain the critical field, $H_C(0)$ = 187.5(5) \si{Oe} (Fig. \ref{Fig1}(i)). The inset of Fig. \ref{Fig1}(h) shows that the five-quadrant magnetization versus magnetic field measurement at 1.8 \si{K} also exhibits a behavior characteristic of type-I superconductors \cite{Pb2Pd,aube}.

The specific heat data shown in Fig. \ref{Fig1}(j) exhibit a superconducting jump at 3.80(3) \si{K}. The low-temperature $C/T$ versus $T^2$ above $T_C$ was well fitted using the Debye-Sommerfeld relation, as shown in the inset (see SM \cite{SM}). The fitting coefficients obtained can be used to calculate the strength of the electron-phonon coupling using the McMillan equation. The value of the electron-phonon coupling constant $\lambda_{e-ph}$, for SnAs, is 0.66(1), indicating a weak strength of electron-phonon coupling. The BCS fitting of the electronic specific heat data reveals that the superconducting gap symmetry exhibits isotropic s-wave behavior (Fig. \ref{Fig1}(j)). Furthermore, the BCS energy gap $\Delta(0)/{k_{B}T_{C}}$ was estimated at 1.79(5) from the best fit data (details in SM \cite{SM}).

\subsection{Electronic Band Structure and Electron-Phonon Coupling}
To analyze the properties of the normal state of SnAs, band structure calculations were performed using density functional theory (DFT) within the generalized gradient approximation (details in the SM \cite{SM}). Fig. \ref{Fig1}(e,f) shows the bulk band structure, the density of states (DOS), and their evolution after the inclusion of the SOC. The non-zero DOS near the Fermi Energy, $E_{F}$, suggests the metallic or semi-metallic character of SnAs. In addition, the manifestation of SOC is strongest at $\Gamma$, producing a maximum band splitting of about 0.45 \si{eV}, breaking the sixfold degeneracy into fourfold. Fig. \textcolor{blue}{S2}(a,b) of the SM \cite{SM} indicates that the states at the Fermi level originate predominantly from hybridized As 4p and Sn 5p orbitals. At the high-symmetry point L, $\sim$1 \si{eV} above $E_{F}$, a Dirac-like crossing is observed, which splits due to SOC. This indicates a possibility of band inversion, which is evident in Fig. \textcolor{blue}{S2}(b) of the SM \cite{SM}, where the spin contributions at the crossing are interchanged. All other splittings (at $\Gamma$, L, and X) occur well away from the Fermi level and do not induce enhancement in the DOS; thus, the role of SOC is less significant in the superconducting properties of SnAs \cite{snasarpes}.

The phonon spectrum calculated using density functional perturbation theory does not host imaginary frequencies, which confirms the dynamical stability of SnAs (SM \cite{SM}, Fig. \textcolor{blue}{S3}(a)). There are six modes in the spectrum (three acoustic and three optical), with the highest optical frequency of approximately 150 \si{cm^{-1}}. The partial atomic phonon DOS (SM \cite{SM}, Fig. \textcolor{blue}{S3}(b)) reveals that higher frequencies are contributed by the As phonon. Assuming conventional electron-phonon pairing within an isotropic Migdal-Eliashberg framework, the mode-resolved spectral function $\alpha^{2}F(\omega)$ and the electron-phonon coupling parameter ($\lambda$) are shown in Fig. \textcolor{blue}{S4} of the SM \cite{SM}, indicating that higher frequency As vibrations play a major role in electron-phonon coupling. The theoretically estimated $\lambda_{e-ph}\approx 0.67$, which gives a superconducting transition temperature $T_{c}\approx3.56$ \si{K} (details in SM \cite{SM}). All these calculations are in good agreement with our experimental values and previous theoretical studies \cite{ElPh}.

\subsection{ZF-\texorpdfstring{$\mu$}{mu}SR}
\begin{figure}[t]
\includegraphics[width=0.9\columnwidth]{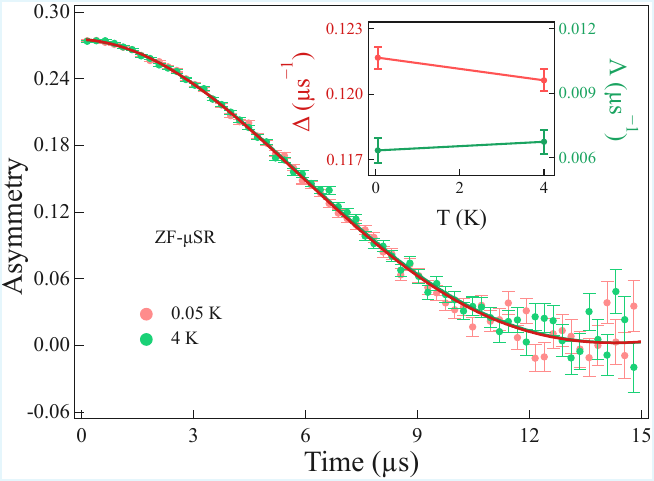}
\caption {\label{fig:2} The ZF-$\mu$SR spectra above and below $T_C$ fitted with an exponentially damped Gaussian KT function. The inset shows the electronic and nuclear relaxation rates, which exhibit negligible changes across the superconducting transition.}
\end{figure}

ZF-$\mu$SR measurements were performed both above $T_C$ (4 \si{K}) and below $T_C$ (0.05 \si{K}) to probe any spontaneous magnetic field present inside the superconducting region that can stem from time-reversal symmetry breaking. The asymmetry spectra do not exhibit any oscillatory component or ordered magnetic structure. Even in the absence of magnetic and electronic moments, the muon asymmetry decays due to the random nuclear moments \cite{Ghosh_2021}. The time evolution of the ZF muon spin polarization is shown in Fig. \ref{fig:2}. The ZF data were fitted using the relaxation function:
\begin{equation}
A(t) = A_{0}G_{ZF}(t)\mathrm{exp}(-\Lambda t)+A_{bg},
\label{eqn:ZF}
\end{equation}
where the coefficients $A_{bg}$ and $A_{0}$ denote the time-independent background asymmetry and the sample asymmetry, respectively. The $\mathrm{exp}(-\Lambda t)$ is the contribution from the rapid fluctuations in the electronic spins and some other relaxations (e.g., time-reversal symmetry breaking), while $G_{ZF}(t)$ stands for the time-dependent Gaussian Kubo-Toyabe (KT) function given by Eq. \ref{Eq:kubo} \cite{hayano1979kubo}. 
\begin{equation}
G_{ZF}(t) = \frac{1}{3}+\frac{2}{3}(1-\Delta^{2}t^{2})\mathrm{exp}\left(\frac{-\Delta^{2}t^{2}}{2}\right).
\label{Eq:kubo}
\end{equation}
The ZF spectra were recorded down to a temperature of 0.05 \si{K}, and no statistically significant differences were observed compared to the normal state spectra. The fitting of both data overlaps and the fitting parameters, i.e., the nuclear relaxation rate $\Delta$ and the electronic relaxation rate $\Lambda$, show minimal variation within the margin of error (see the inset of Fig. \ref{fig:2}). These observations confirm the absence of any spontaneous magnetic field present in the superconducting state. The preserved time-reversal symmetry indicates a single valence state of Sn, as valence fluctuations could lead to an unconventional superconducting ground state, which is easily detectable via ZF-$\mu$SR. The strength of a spontaneous field $|B_{int}|$ can be quantified by $\frac{\sqrt{2}\delta\Delta}{\gamma_{\mu}}$. The difference in nuclear relaxation rate above and below the superconducting region, $\delta\Delta=0.0007(4)$ for SnAs, and the muon gyromagnetic ratio $\gamma_{\mu}$ equals $2\pi\times135.53$ \si{MHz/T}. The calculated magnitude of the spontaneous field is 0.012(3) \si{Oe}, which is much less compared to reports of time-reversal symmetry breaking using the $\mu$SR technique \cite{singh2018re6ti,Sajilesh2024hfrhge,luke1998time}.

\begin{figure}[b]
\includegraphics[width=0.93\columnwidth]{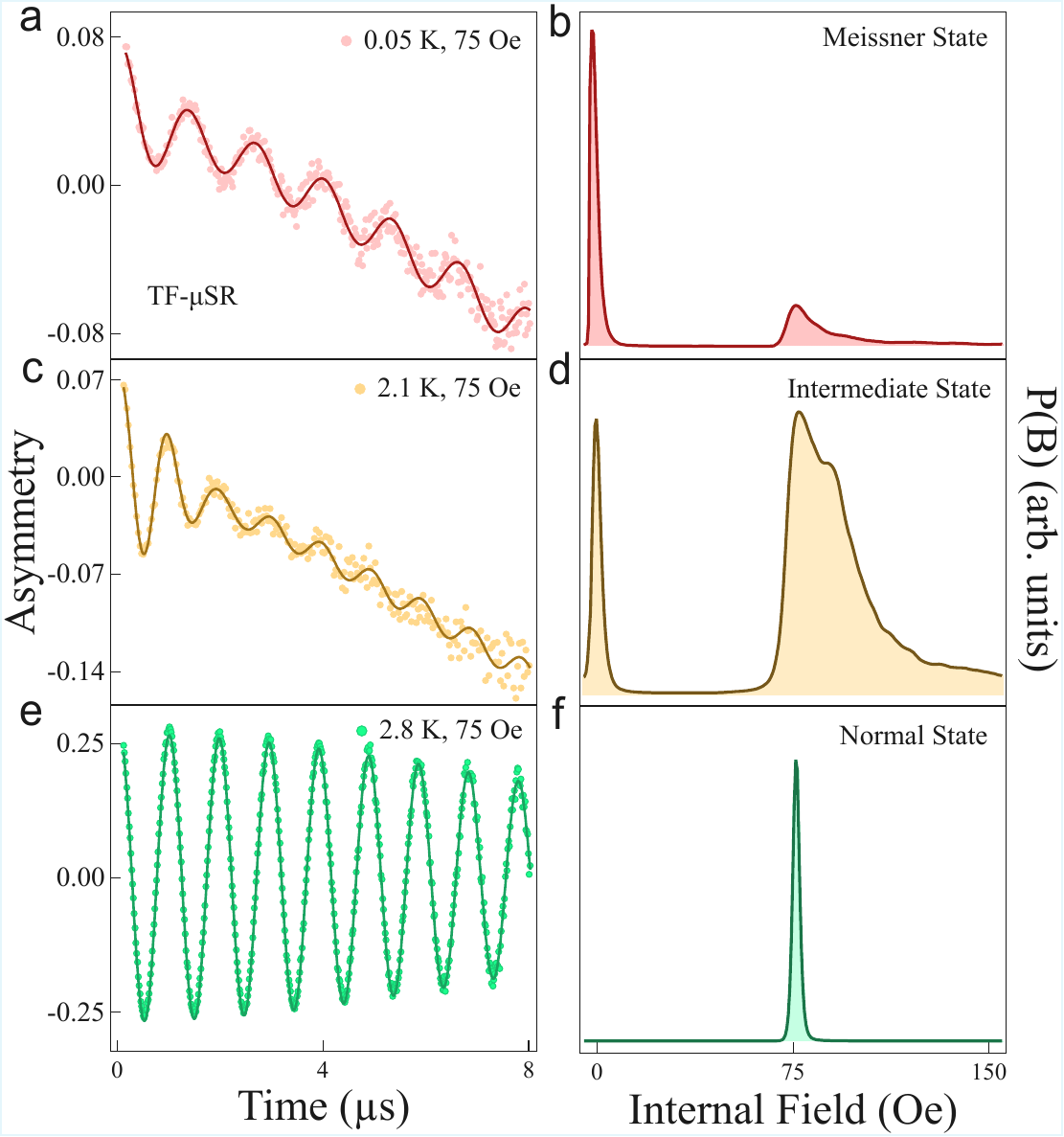}
\caption {\label{Fig:3} The $\mu$SR asymmetry spectra in TF configuration for (a) 0.05 \si{K}, (c) 2.1 \si{K}, and (e) 2.8 \si{K} at a magnetic field of 75 \si{Oe}, fitted using Eq. \ref{Eq:TF}. The field distribution of the local field obtained from the MaxEnt transformation of the corresponding $\mu$SR spectra, denoting (b) Meissner, (d) intermediate, and (f) normal states.}
\end{figure}

\subsection{TF-\texorpdfstring{$\mu$}{mu}SR}
Type-I superconductors typically exhibit three distinct states: the Meissner state, the intermediate state, and the normal state. These three states can be differentiated by analyzing the internal field distribution using the TF-$\mu$SR technique and the Maximum Entropy (MaxEnt) spectra. MaxEnt analysis employs an iteration-based method to obtain a frequency spectrum where entropy is maximized based on asymmetry data \cite{maxent}. The asymmetry spectra depend on the superconducting state of the sample, and the volume fraction of the particular state determines the probability of the muons encountering the internal field.

\begin{figure}[t]
\includegraphics[width=0.95\columnwidth]{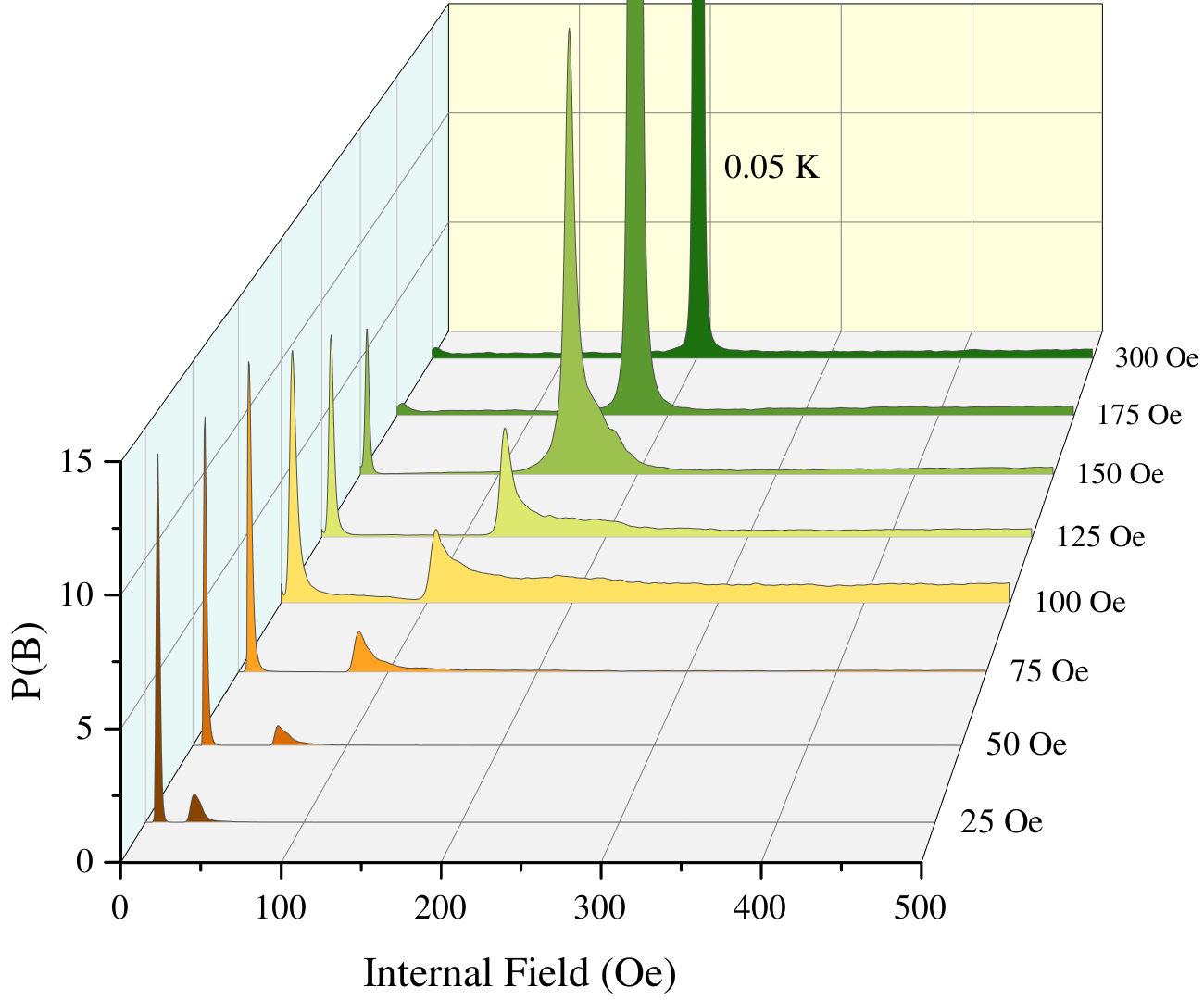}
\caption {\label{Fig:4} The internal field distribution P(B), at 0.05 \si{K} under varying magnetic fields up to 300 \si{Oe}.}
\end{figure}

\begin{figure*}
\includegraphics[width=2.1\columnwidth]{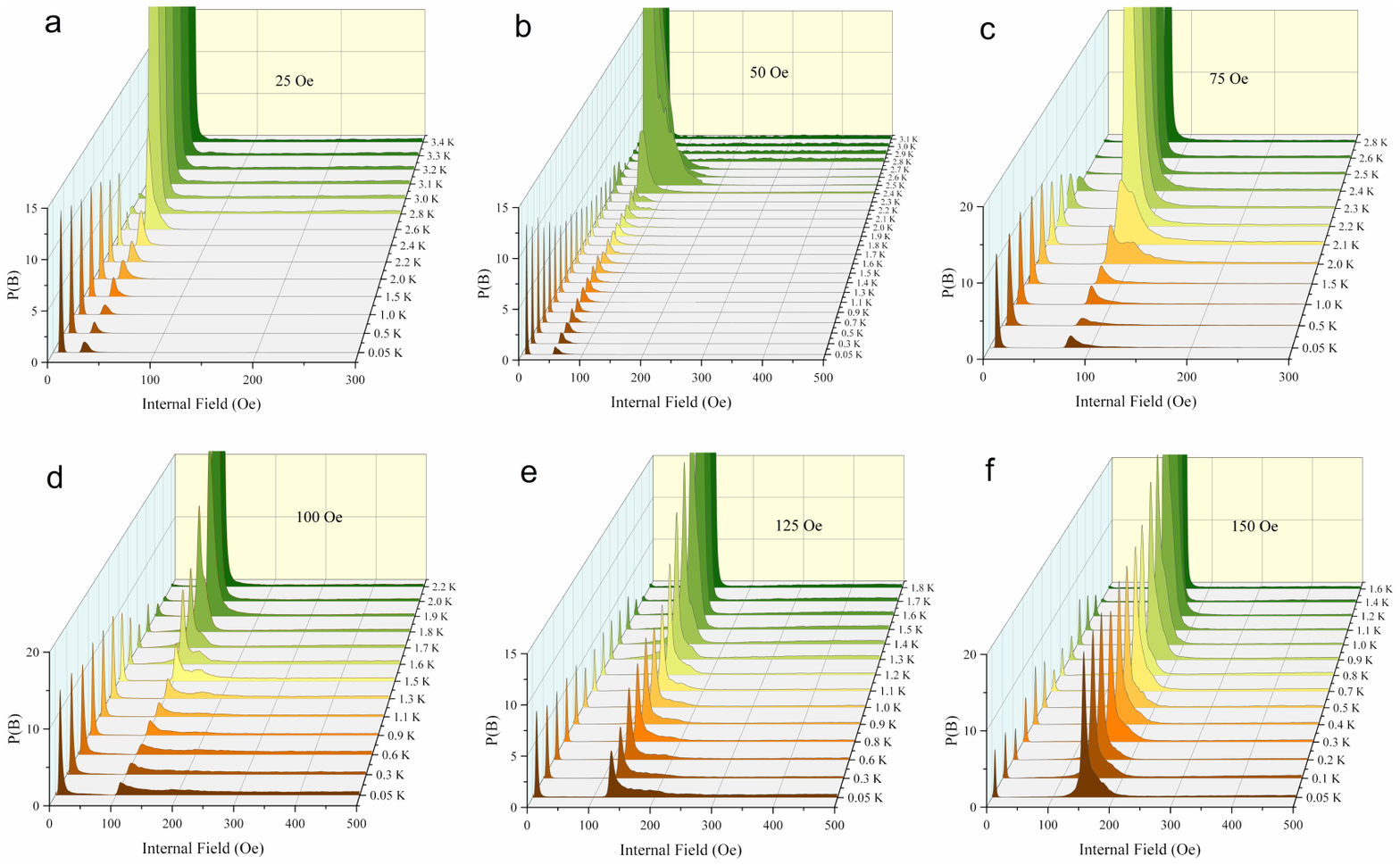}
\caption {\label{fig:5} The temperature-dependent field distribution of the internal field at constant magnetic fields between 25 \si{Oe} and 150 \si{Oe}. (a-d) In a particular field between 25 \si{Oe} to 100 \si{Oe}, the 3D plots show the transition from the Meissner state to the intermediate state, followed by the normal state, as the temperature increases. (e,f) At 125 \si{Oe} and 150 \si{Oe}, there is a transition from the intermediate state to the normal state.}
\end{figure*}

The transverse field data were taken in the field-cooled mode, down to the lowest accessible temperature of 0.05 \si{K}, for a range of applied magnetic fields between 25 and 300 \si{Oe}. Fig. \ref{Fig:3} exemplifies the asymmetry spectra and their MaxEnt transformation in the Meissner, intermediate, and normal states under an applied magnetic field of 75 \si{Oe}. The asymmetry spectra at 0.05 \si{K} exhibit slow KT-relaxation due to the inclusion of static nuclear moments in the Meissner state, depicted in Fig. \ref{Fig:3}(a). The corresponding MaxEnt spectra hosts a sharp peak at zero magnetic field, giving a clear signature of the Meissner state, and the weak contribution at 75 \si{Oe} comes from the background signals of muons interacting with the instrument parts, such as the sample holder (Fig. \ref{Fig:3}(b)). Due to the presence of the demagnetization factor, the intermediate state allows the normal region inside the sample to experience a field that is greater than the applied field, equal to the critical field. Hence, Fig. \ref{Fig:3}(d) shows the three peaks in the intermediate state, which are observed at zero magnetic field, the applied field, and a higher field ($\approx90$ \si{Oe}), confirming the existence of regions with magnetic flux expulsion. The detection of an intermediate state strongly affirms type-I superconductivity in SnAs single crystals. For type-II superconductors, we expect a peak at a field lower than the applied field, which is caused by the formation of a flux line lattice (FLL) in the mixed state. In the asymmetry spectra for the intermediate and normal states, the Gaussian-relaxed sinusoidal oscillatory function is included for non-zero magnetic fields (refer to Fig. \ref{Fig:3}(c,e)). At 2.8 \si{K}, the superconductor is in the normal state and the asymmetry spectra have a homogeneous field distribution without KT relaxation. The MaxEnt spectra in the normal state have only one peak at the applied magnetic field, validating the penetration of the magnetic field in the normal region, as shown in Fig. \ref{Fig:3}(f). The function used to fit the TF muon asymmetry data is given by \cite{aube,maisuradze2009comparison,weber1993flux}:

\begin{equation}
G(t) = A(t)+\sum_{i=1}^N A_{i}\exp\left(-\frac{1}{2}\sigma_i^2t^2\right)\cos(\gamma_\mu B_it+\phi),
\label{Eq:TF}
\end{equation}

where $A(t)$ refers to Eq. \ref{eqn:ZF}, $A_{i}$ and $\phi$ are the initial asymmetry and phase offset, $\sigma_{i}$ and $B_i$ are the $i^{th}$ components of the Gaussian relaxation rate and field distribution, respectively. The best fittings for the asymmetry spectra were described in $N=2$, and $\sigma_{1} = 0$ is fixed.

The internal field distribution, P(B), at a constant temperature of 0.05 \si{K} for various applied magnetic fields from 25 \si{Oe} to 300 \si{Oe} was obtained from the MaxEnt transform of the asymmetry data, as shown in Fig. \ref{Fig:4}. At a temperature of 0.05 \si{K}, SnAs remains in the Meissner state at 25 \si{Oe}, transitions to the intermediate state for applied fields greater than 125 \si{Oe} and finally enters the normal state at 175 \si{Oe}. When the sample is in the intermediate state, there is a peak in the magnetic field, $H_C$, corresponding to the critical magnetic field, which persists in varying the applied field, assuring the superconducting type-I nature. The additional peak in type-II superconductors remains below the applied field and relocates as the magnetic field increases. The intensity of the peak near zero field (the applied field) increases (decreases) coherently. The intensity of the zero-field peak clearly shows maxima at the lowest field in the Meissner state. As the Meissner volume fraction decreases while moving towards the intermediate state, the peak intensity decreases and finally vanishes upon entering the normal state. 

Furthermore, the temperature variation of P(B) in various applied fields between 25 \si{Oe} and 150 \si{Oe} at 25 \si{Oe} intervals is demonstrated in Fig. \ref{fig:5}. These measurements were executed in the field-cooled warming mode for each magnetic field. All 3D plots (Fig. \ref{fig:5}(a-d)) show the decrease (increase) in the intensity of the peak at zero-field (applied field) when moving from the Meissner state to the normal state via the intermediate state. A similar intensity change is observed in Fig. \ref{fig:5}(e,f) while evolving from the intermediate to the normal state with field. For each magnetic field, it is possible to identify the temperature range in which the sample is categorized as either the Meissner state, the intermediate state, or the normal state. 

\begin{figure}[t]
\includegraphics[width=0.95\columnwidth]{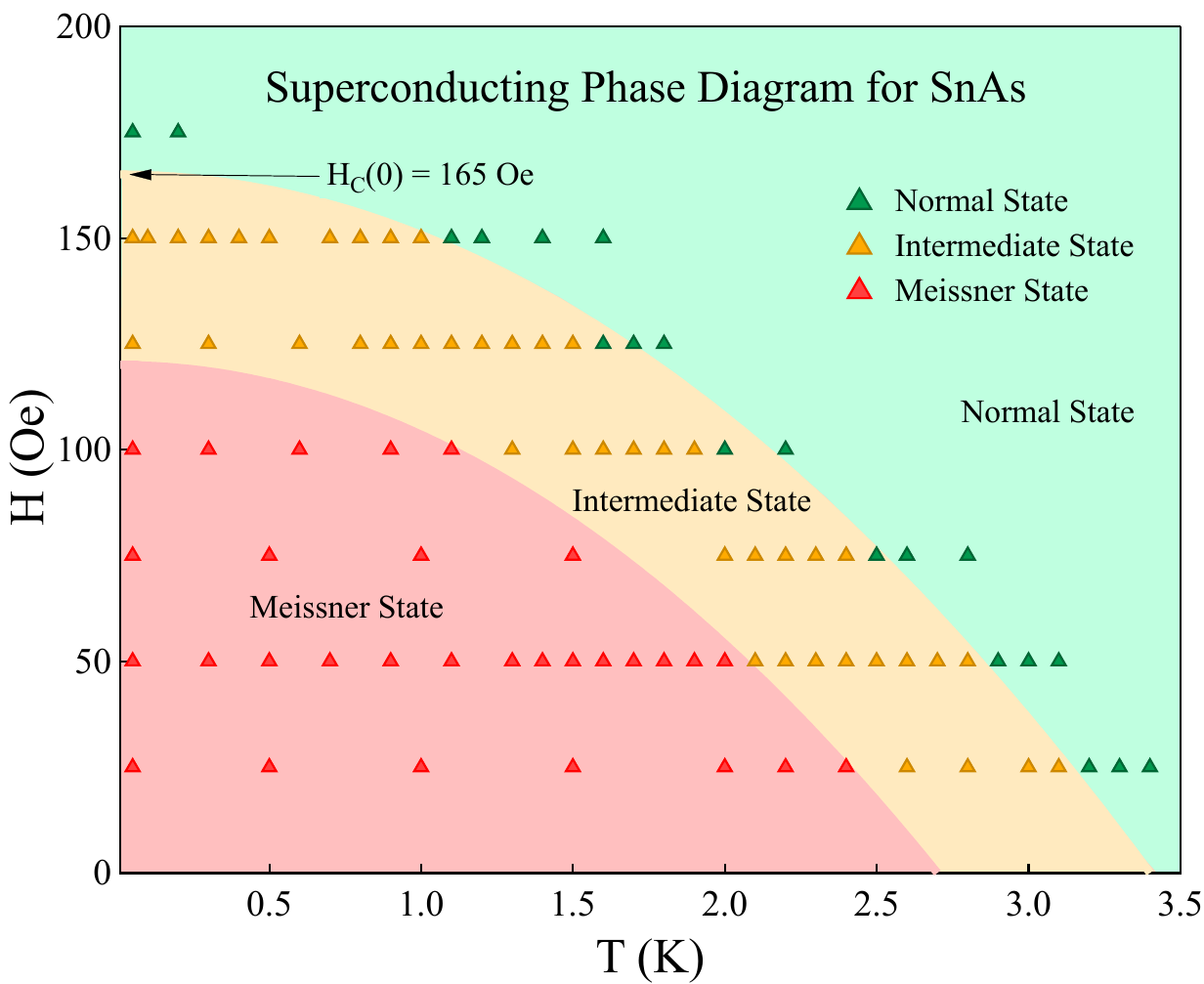}
\caption {\label{Fig:6} Superconducting phase diagram of SnAs, indicating the approximate region for the Meissner, intermediate, and normal states, depending on the superconducting nature at each data point in the H-T plane. The color of the data points indicates their superconducting behavior, as determined by the MaxEnt analysis of the field distribution probed by muons, as shown in Fig. \ref{fig:5}.}
\end{figure}

\subsection{Superconducting Phase Diagram}
The comprehensive TF-$\mu$SR measurements were performed over a wide range of temperatures and magnetic fields to map the superconducting phase diagram of SnAs. Based on the local field distribution P(B) described in Fig. \ref{fig:5}, the position of the data set in the phase diagram was qualitatively classified into the Meissner, intermediate or normal region. Fig. \ref{Fig:6} shows the superconducting phase diagram of SnAs, where the red, yellow, and green regions correspond to the Meissner, intermediate, and normal states, respectively. The red, yellow, and green triangles are the analogous data points in the H-T plane for each distribution at a fixed applied magnetic field and temperature in Fig. \ref{fig:5}. The lines separating the Meissner-intermediate region and the intermediate-normal region were determined on the basis of the superconducting state at each data point. It was observed that the G-L equation (SM \cite{SM}, Eq. \textcolor{blue}{1}) properly separates the Meissner, intermediate, and normal regions. Only a few normal-state points at the boundary of the intermediate-normal region overlapped with the intermediate region, which is within the error limit of the measurement conditions. The superconducting critical field $H_C$ was found to be 165(1) \si{Oe} based on the superconducting phase diagram. This phase diagram establishes type-I superconductivity in single crystals of SnAs through a microscopic investigation using $\mu$SR, with a precise determination of the intermediate state.

\section{Summary and Conclusions}
This work presents a detailed $\mu$SR analysis of the topological candidate superconductor SnAs. The single crystals were synthesized using the modified Bridgman method. It is characterized by X-ray diffraction and magnetic and heat capacity measurements. The Zero-field $\mu$SR confirms the preservation of time-reversal symmetry, confirming the absence of an unconventional superconducting state. This observation suggests that valence-fluctuations may not be present in SnAs, which typically leads to an unconventional ground state via negative-$U$ potentials \cite{AgSnSe2,valencetaraphder,valencevarma}. TF-$\mu$SR measurements were performed at various magnetic fields and temperatures, revealing the type-I nature of superconductivity in the SnAs compound, with clear signatures of the Meissner, intermediate and normal states. It excludes any possibility for mixed Type-I/Type-II behavior. We further present the superconducting phase diagram, which indicates the well-defined regions of the three states, quantitatively separated by the G-L equation. Furthermore, the similarity between the theoretically calculated and experimentally observed $\lambda_{e-ph}$ and $T_{C}$ provides strong evidence that weak electron-phonon pairing is responsible for superconductivity in SnAs. The coexistence of nontrivial band topology and strong spin–orbit coupling in this type-I superconductor with preserved time-reversal symmetry suggests the potential realization of topological superconductivity in SnAs \cite{TSCinTM}, which warrants further theoretical and experimental exploration.

\section{Acknowledgments}
S.~S. acknowledges the funding agency, University Grants Commission (UGC), Government of India, for the Senior Research Fellowship (SRF). R.~P.~S. acknowledges the ANRF Government of India for the Core Research Grant No. CRG/2023/000817 and ISIS, STFC, UK, for providing beamtime for the $\mu$SR experiments.

\nocite{*}
\bibliographystyle{revtex}
%\printbibliography
\bibliography{Library}

@article{valencetaraphder,
  title = {Heavy-fermion behavior in a negative-U Anderson model},
  author = {Taraphder, A. and Coleman, P.},
  journal = {Phys. Rev. Lett.},
  volume = {66},
  issue = {21},
  pages = {2814--2817},
  numpages = {0},
  year = {1991},
  month = {May},
  publisher = {American Physical Society},
  doi = {10.1103/PhysRevLett.66.2814},
  url = {https://link.aps.org/doi/10.1103/PhysRevLett.66.2814}
}

@article{valencevarma,
  title = {Missing valence states, diamagnetic insulators, and superconductors},
  author = {Varma, C. M.},
  journal = {Phys. Rev. Lett.},
  volume = {61},
  issue = {23},
  pages = {2713--2716},
  numpages = {0},
  year = {1988},
  month = {Dec},
  publisher = {American Physical Society},
  doi = {10.1103/PhysRevLett.61.2713},
  url = {https://link.aps.org/doi/10.1103/PhysRevLett.61.2713}
}

@article{AgSnSe2,
  title = {Superconducting ground state study of the valence-skipped compound ${\mathrm{AgSnSe}}_{2}$},
  author = {Kataria, A. and Arushi and Sharma, S. and Agarwal, T. and Pula, M. and Beare, J. and Yoon, S. and Cai, Y. and Kojima, K. M. and Luke, G. M. and Singh, R. P.},
  journal = {Phys. Rev. B},
  volume = {107},
  issue = {17},
  pages = {174517},
  numpages = {7},
  year = {2023},
  month = {May},
  publisher = {American Physical Society},
  doi = {10.1103/PhysRevB.107.174517},
  url = {https://link.aps.org/doi/10.1103/PhysRevB.107.174517}
}

@article{ElPh,
title = {Electron–phonon interaction and superconductivity in SnAs with the sodium chloride crystal structure},
journal = {Solid State Commun.},
volume = {221},
pages = {24-27},
year = {2015},
issn = {0038-1098},
doi = {https://doi.org/10.1016/j.ssc.2015.08.006},
url = {https://www.sciencedirect.com/science/article/pii/S0038109815002896},
author = {H. M. Tütüncü and G. P. Srivastava}
}

@article{cuxbi2se3,
  title = {Odd-Parity Topological Superconductors: Theory and Application to ${\mathrm{Cu}}_{x}{\mathrm{Bi}}_{2}{\mathrm{Se}}_{3}$},
  author = {Fu, Liang and Berg, Erez},
  journal = {Phys. Rev. Lett.},
  volume = {105},
  issue = {9},
  pages = {097001},
  numpages = {4},
  year = {2010},
  month = {Aug},
  publisher = {American Physical Society},
  doi = {10.1103/PhysRevLett.105.097001},
  url = {https://link.aps.org/doi/10.1103/PhysRevLett.105.097001}
}

@article{TSCinTM,
author = {Li, Yupeng and Xu, Zhu-An},
title = {Exploring Topological Superconductivity in Topological Materials},
journal = {Adv. Quantum Technol.},
volume = {2},
pages = {1800112},
doi = {https://doi.org/10.1002/qute.201800112},
url = {https://advanced.onlinelibrary.wiley.com/doi/abs/10.1002/qute.201800112},
year = {2019}
}

@article{topomat,
   author = "Yan, Binghai and Felser, Claudia",
   title = "Topological Materials: Weyl Semimetals", 
   journal= "Annu. Rev. Condens. Matter Phys.",
   year = "2017",
   volume = "8",
   number = "Volume 8, 2017",
   pages = "337-354",
   doi = "https://doi.org/10.1146/annurev-conmatphys-031016-025458",
   url = "https://www.annualreviews.org/content/journals/10.1146/annurev-conmatphys-031016-025458",
   publisher = "Annual Reviews"
}

@article{viewpoint,
author = {Kumar, Nitesh and Guin, Satya N. and Manna, Kaustuv and Shekhar, Chandra and Felser, Claudia},
title = {Topological Quantum Materials from the Viewpoint of Chemistry},
journal = {Chem. Rev.},
volume = {121},
number = {5},
pages = {2780-2815},
year = {2021},
doi = {10.1021/acs.chemrev.0c00732},
URL = {https://doi.org/10.1021/acs.chemrev.0c00732}
}

@article{pdte2,
  title = {Type-I superconductivity in the Dirac semimetal ${\mathrm{PdTe}}_{2}$},
  author = {Leng, H. and Paulsen, C. and Huang, Y. K. and de Visser, A.},
  journal = {Phys. Rev. B},
  volume = {96},
  issue = {22},
  pages = {220506},
  numpages = {5},
  year = {2017},
  month = {Dec},
  publisher = {American Physical Society},
  doi = {10.1103/PhysRevB.96.220506},
  url = {https://link.aps.org/doi/10.1103/PhysRevB.96.220506}
}

@article{al6re,
  title = {Type-I superconductivity in ${\mathrm{Al}}_{6}\mathrm{Re}$},
  author = {Peets, Darren C. and Cheng, Erjian and Ying, Tianping and Kriener, Markus and Shen, Xiaoping and Li, Shiyan and Feng, Donglai},
  journal = {Phys. Rev. B},
  volume = {99},
  issue = {14},
  pages = {144519},
  numpages = {9},
  year = {2019},
  month = {Apr},
  publisher = {American Physical Society},
  doi = {10.1103/PhysRevB.99.144519},
  url = {https://link.aps.org/doi/10.1103/PhysRevB.99.144519}
}

@Article{irsn4dft,
author ="Mai, Thi Ly and Tran, Vinh Hung",
title  ="Ab initio investigation of electronic structure and optical properties of IrSn4",
journal  ="RSC Adv.",
year  ="2022",
volume  ="12",
issue  ="28",
pages  ="17882-17888",
publisher  ="The Royal Society of Chemistry",
doi  ="10.1039/D2RA01672A",
url  ="http://dx.doi.org/10.1039/D2RA01672A"
}

@article{irsn4,
author = {Ahmad ,Nazir and Shimada ,Shunsuke and Hasegawa ,Takumi and Suzuki ,Hiroto and Afzal ,Md Asif and Nakamura ,Naoki and Higashinaka ,Ryuji and Matsuda ,Tatsuma D. and Aoki ,Yuji},
title = {Linear Magnetoresistance and Type-I Superconductivity in \(\beta\)-IrSn4},
journal = {J. Phys. Soc. Jpn.},
volume = {93},
number = {4},
pages = {044706},
year = {2024},
doi = {10.7566/JPSJ.93.044706},
URL ={https://doi.org/10.7566/JPSJ.93.044706}
}

@article{lipd2ge,
  title = {Soft-mode enhanced type-I superconductivity in $\mathrm{Li}{\mathrm{Pd}}_{2}\mathrm{Ge}$},
  author = {G\'ornicka, Karolina and Kuderowicz, Gabriel and Carnicom, Elizabeth M. and Kutorasi\ifmmode \acute{n}\else \'{n}\fi{}ski, Kamil and Wiendlocha, Bartlomiej and Cava, Robert J. and Klimczuk, Tomasz},
  journal = {Phys. Rev. B},
  volume = {102},
  issue = {2},
  pages = {024507},
  numpages = {13},
  year = {2020},
  month = {Jul},
  publisher = {American Physical Society},
  doi = {10.1103/PhysRevB.102.024507},
  url = {https://link.aps.org/doi/10.1103/PhysRevB.102.024507}
}

@article{pdte2type1,
  title = {Type-I superconductivity in the Dirac semimetal ${\mathrm{PdTe}}_{2}$ probed by $\ensuremath{\mu}\mathrm{SR}$},
  author = {Leng, H. and Orain, J.-C. and Amato, A. and Huang, Y. K. and de Visser, A.},
  journal = {Phys. Rev. B},
  volume = {100},
  issue = {22},
  pages = {224501},
  numpages = {8},
  year = {2019},
  month = {Dec},
  publisher = {American Physical Society},
  doi = {10.1103/PhysRevB.100.224501},
  url = {https://link.aps.org/doi/10.1103/PhysRevB.100.224501}
}

@article{Sntype1,
  title = {Muon spin rotation study of type-I superconductivity: Elemental $\ensuremath{\beta}$-Sn},
  author = {Karl, Richard and Burri, Florence and Amato, Alex and Doneg\`a, Mauro and Gvasaliya, Severian and Luetkens, Hubertus and Morenzoni, Elvezio and Khasanov, Rustem},
  journal = {Phys. Rev. B},
  volume = {99},
  issue = {18},
  pages = {184515},
  numpages = {13},
  year = {2019},
  month = {May},
  publisher = {American Physical Society},
  doi = {10.1103/PhysRevB.99.184515},
  url = {https://link.aps.org/doi/10.1103/PhysRevB.99.184515}
}

@Article{LaCuSb2,
author ="Lygouras, Chris J. and Zhang, Junyi and Gautreau, Jonah and Pula, Mathew and Sharma, Sudarshan and Gao, Shiyuan and Berry, Tanya and Halloran, Thomas and Orban, Peter and Grissonnanche, Gael and Chamorro, Juan R. and Mikuri, Taketora and Bhoi, Dilip K. and Siegler, Maxime A. and Livi, Kenneth J.T. and Uwatoko, Yoshiya and Nakatsuji, Satoru and Ramshaw, B. J. and Li, Yi and Luke, Graeme M. and Broholm, Collin L. and McQueen, Tyrel M.",
title  ="Type I and type II superconductivity in a quasi-2D Dirac metal",
journal  ="Mater. Adv.",
year  ="2025",
volume  ="6",
issue  ="5",
pages  ="1685-1694",
publisher  ="RSC",
doi  ="10.1039/D5MA00022J",
url  ="http://dx.doi.org/10.1039/D5MA00022J"
}

@article{nbge2coexistence,
  title = {$\ensuremath{\mu}\mathrm{SR}$ study on the noncentrosymmetric superconductor ${\mathrm{NbGe}}_{2}$},
  author = {Jiao, J. C. and Chen, K. W. and Hillier, Adrian D. and Ito, T. U. and Higemoto, W. and Li, Zheng and Lv, Baijiang and Xu, Zhu-An and Shu, Lei},
  journal = {Phys. Rev. B},
  volume = {110},
  issue = {21},
  pages = {214516},
  numpages = {9},
  year = {2024},
  month = {Dec},
  publisher = {American Physical Society},
  doi = {10.1103/PhysRevB.110.214516},
  url = {https://link.aps.org/doi/10.1103/PhysRevB.110.214516}
}

@article{zrb12coexistence,
  title = {Coexistence of type-I and type-II superconductivity signatures in $\mathrm{Zr}{\mathrm{B}}_{12}$ probed by muon spin rotation measurements},
  author = {Biswas, P. K. and Rybakov, F. N. and Singh, R. P. and Mukherjee, Saumya and Parzyk, N. and Balakrishnan, G. and Lees, M. R. and Dewhurst, C. D. and Babaev, E. and Hillier, A. D. and Paul, D. Mc K.},
  journal = {Phys. Rev. B},
  volume = {102},
  issue = {14},
  pages = {144523},
  numpages = {6},
  year = {2020},
  month = {Oct},
  publisher = {American Physical Society},
  doi = {10.1103/PhysRevB.102.144523},
  url = {https://link.aps.org/doi/10.1103/PhysRevB.102.144523}
}

@article{pdte2coexistence,
    title={Coexistence of type-I and type-II superconductivity in topological superconductor PdTe $ \_2$}, 
    author={Singh, D and Biswas, Pabitra K and Yoon, Sungwon and Lee, C H and Hillier, A D and Singh, R P and Singh, Amit Y and Choi, K-Y},
    journal={arXiv:1910.13773},
    url={https://arxiv.org/abs/1910.13773}, 
    year={2019}
}

@article{snasheatcap,
doi = {10.1088/1361-648X/ac6474},
url = {https://dx.doi.org/10.1088/1361-648X/ac6474},
year = {2022},
month = {apr},
publisher = {IOP Publishing},
volume = {34},
number = {25},
pages = {255702},
author = {Sharma, M M and Awana, V P S},
title = {Detailed magneto heat capacity analysis of SnAs topological superconductor},
journal = {J. Phys.: Condens. Matter}
}

@article{sntetopological,
  title={Topological crystalline insulators in the SnTe material class},
  author={Hsieh, Timothy H and Lin, Hsin and Liu, Junwei and Duan, Wenhui and Bansil, Arun and Fu, Liang},
  journal={Nat. Commun.},
  volume={3},
  number={1},
  pages={982},
  year={2012},
  doi={https://doi.org/10.1038/ncomms1969},
  publisher={Nature Publishing Group UK London}
}

@article{SnSeSnS,
  title = {Rocksalt SnS and SnSe: Native topological crystalline insulators},
  author = {Sun, Yan and Zhong, Zhicheng and Shirakawa, Tomonori and Franchini, Cesare and Li, Dianzhong and Li, Yiyi and Yunoki, Seiji and Chen, Xing-Qiu},
  journal = {Phys. Rev. B},
  volume = {88},
  issue = {23},
  pages = {235122},
  numpages = {6},
  year = {2013},
  month = {Dec},
  publisher = {American Physical Society},
  doi = {10.1103/PhysRevB.88.235122},
  url = {https://link.aps.org/doi/10.1103/PhysRevB.88.235122}
}

@article{SnP,
  title = {Superconductivity at the Polar-Nonpolar Phase Boundary of SnP with an Unusual Valence State},
  author = {Kamitani, M. and Bahramy, M. S. and Nakajima, T. and Terakura, C. and Hashizume, D. and Arima, T. and Tokura, Y.},
  journal = {Phys. Rev. Lett.},
  volume = {119},
  issue = {20},
  pages = {207001},
  numpages = {6},
  year = {2017},
  month = {Nov},
  publisher = {American Physical Society},
  doi = {10.1103/PhysRevLett.119.207001},
  url = {https://link.aps.org/doi/10.1103/PhysRevLett.119.207001}
}

@article{mixedvalence,
title = {Valency, valence degeneracy, ferroelectricity, and superconductivity},
journal = {Prog. Solid State Chem.},
volume = {37},
number = {4},
pages = {251-261},
year = {2009},
issn = {0079-6786},
doi = {https://doi.org/10.1016/j.progsolidstchem.2010.08.001},
url = {https://www.sciencedirect.com/science/article/pii/S0079678610000026},
author = {Arthur W. Sleight}
}

@article{naclprl,
  title = {Superconductivity of Intermetallic Compounds with NaCl-Type and Related Structures},
  author = {Geller, S. and Hull, G. W.},
  journal = {Phys. Rev. Lett.},
  volume = {13},
  issue = {4},
  pages = {127--129},
  numpages = {0},
  year = {1964},
  month = {Jul},
  publisher = {American Physical Society},
  doi = {10.1103/PhysRevLett.13.127},
  url = {https://link.aps.org/doi/10.1103/PhysRevLett.13.127}
}

@article{sn0.4sb0.6,
title = {Type-II superconductivity below 4K in Sn0.4Sb0.6},
journal = {J. Alloys Compd.},
volume = {844},
pages = {156140},
year = {2020},
issn = {0925-8388},
doi = {https://doi.org/10.1016/j.jallcom.2020.156140},
url = {https://www.sciencedirect.com/science/article/pii/S0925838820325044},
author = {M. M. Sharma and Kapil Kumar and Lina Sang and X. L. Wang and V. P. S. Awana}
}

@article{snasarpes,
  title={ARPES measurements of SnAs electronic band structure},
  author={Bezotosnyi, Pavel Igorevich and Dmitrieva, Kristina Alekseevna and Gavrilkin, S Yu and Pervakov, Kirill Sergeevich and Tsvetkov, A Yu and Martovitski, VP and Rybkin, Artem Gennadievich and Vilkov, O Yu and Pudalov, Vladimir Moiseevich},
  journal={JETP Lett.},
  volume={106},
  number={8},
  pages={514--516},
  year={2017},
  doi={https://doi.org/10.1134/S0021364017200024},
  publisher={Springer}
}

@article{snsbx,
  title = {Superconducting phase diagram and nontrivial band topology of structurally modulated ${\mathrm{Sn}}_{1\ensuremath{-}x}{\mathrm{Sb}}_{x}$},
  author = {Liu, Bin and Xiao, Chengcheng and Zhu, Qinqing and Wu, Jifeng and Cui, Yanwei and Wang, Hangdong and Wang, Zhicheng and Lu, Yunhao and Ren, Zhi and Cao, Guang-han},
  journal = {Phys. Rev. Mater.},
  volume = {3},
  issue = {8},
  pages = {084603},
  numpages = {7},
  year = {2019},
  month = {Aug},
  publisher = {American Physical Society},
  doi = {10.1103/PhysRevMaterials.3.084603},
  url = {https://link.aps.org/doi/10.1103/PhysRevMaterials.3.084603}
}

@article{snte,
  title={Experimental realization of a topological crystalline insulator in SnTe},
  author={Tanaka, Y and Ren, Zhi and Sato, T and Nakayama, K and Souma, S and Takahashi, T and Segawa, Kouji and Ando, Yoichi},
  journal={Nat. Phys.},
  volume={8},
  number={11},
  pages={800--803},
  year={2012},
  doi={https://doi.org/10.1038/nphys2442},
  publisher={Nature Publishing Group UK London}
}

@article{maxent,
  title={$\mu$SR frequency spectra using the maximum entropy method},
  author={Rainford, B D and Daniell, G J},
  journal={Hyperfine Interact.},
  volume={87},
  number={1},
  pages={1129--1134},
  year={1994},
  doi={https://doi.org/10.1007/BF02068515},
  publisher={Springer}
}

@article{snastypeII,
title = {SnAs: A 4K weak type-II superconductor with non-trivial band topology},
journal = {Solid State Commun.},
volume = {340},
pages = {114531},
year = {2021},
issn = {0038-1098},
doi = {https://doi.org/10.1016/j.ssc.2021.114531},
url = {https://www.sciencedirect.com/science/article/pii/S0038109821003227},
author = {M. M. Sharma and N. K. Karn and Prince Sharma and Ganesh Gurjar and S. Patnaik and V. P. S. Awana}
}

@article{snassheet,
  title={Andreev reflection spectroscopy on SnAs single crystals},
  author={Howlader, Sandeep and Mehta, Nikhlesh Singh and Sharma, M M and Awana, V P S and Sheet, Goutam},
  journal={J. Supercond. Novel Magn.},
  volume={35},
  number={7},
  pages={1839--1845},
  year={2022},
  doi={https://doi.org/10.1007/s10948-022-06261-1},
  publisher={Springer}
}

@article{snasvalence,
  title={SnAs with the NaCl-type structure: Type-I superconductivity and single valence state of Sn},
  author={Wang, Yue and Sato, Hikaru and Toda, Yoshitake and Ueda, Shigenori and Hiramatsu, Hidenori and Hosono, Hideo},
  journal={Chem. Mater.},
  volume={26},
  number={24},
  pages={7209--7213},
  year={2014},
  doi={dx.doi.org/10.1021/cm503992d},
  publisher={ACS Publications}
}

@article{snasprb,
  title = {Electronic band structure and superconducting properties of SnAs},
  author = {Bezotosnyi, P. I. and Dmitrieva, K. A. and Sadakov, A. V. and Pervakov, K. S. and Muratov, A. V. and Usoltsev, A. S. and Tsvetkov, A. Yu. and Gavrilkin, S. Yu. and Pavlov, N. S. and Slobodchikov, A. A. and Vilkov, O. Yu. and Rybkin, A. G. and Nekrasov, I. A. and Pudalov, V. M.},
  journal = {Phys. Rev. B},
  volume = {100},
  issue = {18},
  pages = {184514},
  numpages = {12},
  year = {2019},
  month = {Nov},
  publisher = {American Physical Society},
  doi = {10.1103/PhysRevB.100.184514},
  url = {https://link.aps.org/doi/10.1103/PhysRevB.100.184514}
}

@article{aube,
  title = {Type-I superconductivity in the noncentrosymmetric superconductor BeAu},
  author = {Singh, D. and Hillier, A. D. and Singh, R. P.},
  journal = {Phys. Rev. B},
  volume = {99},
  issue = {13},
  pages = {134509},
  numpages = {9},
  year = {2019},
  month = {Apr},
  publisher = {American Physical Society},
  doi = {10.1103/PhysRevB.99.134509},
  url = {https://link.aps.org/doi/10.1103/PhysRevB.99.134509}
}

@article{liu2014prx,
  title = {Non-Abelian Majorana Doublets in Time-Reversal-Invariant Topological Superconductors},
  author = {Liu, Xiong-Jun and Wong, Chris L. M. and Law, K. T.},
  journal = {Phys. Rev. X},
  volume = {4},
  issue = {2},
  pages = {021018},
  numpages = {16},
  year = {2014},
  month = {Apr},
  publisher = {American Physical Society},
  doi = {10.1103/PhysRevX.4.021018},
  url = {https://link.aps.org/doi/10.1103/PhysRevX.4.021018}
}

@article{wang2018tsm,
    author = {Wang, Jian},
    title = {Superconductivity in topological semimetals},
    journal = {Natl. Sci. Rev.},
    volume = {6},
    number = {2},
    pages = {199-202},
    year = {2018},
    month = {12},
    issn = {2095-5138},
    doi = {10.1093/nsr/nwy155},
    url = {https://doi.org/10.1093/nsr/nwy155}
}

@article{qi2011tis,
  title = {Topological insulators and superconductors},
  author = {Qi, Xiao-Liang and Zhang, Shou-Cheng},
  journal = {Rev. Mod. Phys.},
  volume = {83},
  issue = {4},
  pages = {1057--1110},
  numpages = {0},
  year = {2011},
  month = {Oct},
  publisher = {American Physical Society},
  doi = {10.1103/RevModPhys.83.1057},
  url = {https://link.aps.org/doi/10.1103/RevModPhys.83.1057}
}

@article{nayak2008topo,
  title = {Non-Abelian anyons and topological quantum computation},
  author = {Nayak, Chetan and Simon, Steven H. and Stern, Ady and Freedman, Michael and Das Sarma, Sankar},
  journal = {Rev. Mod. Phys.},
  volume = {80},
  issue = {3},
  pages = {1083--1159},
  numpages = {0},
  year = {2008},
  month = {Sep},
  publisher = {American Physical Society},
  doi = {10.1103/RevModPhys.80.1083},
  url = {https://link.aps.org/doi/10.1103/RevModPhys.80.1083}
}

@article{Sato,
doi = {10.1088/1361-6633/aa6ac7},
url = {https://dx.doi.org/10.1088/1361-6633/aa6ac7},
year = {2017},
month = {may},
publisher = {IOP Publishing},
volume = {80},
number = {7},
pages = {076501},
author = {Sato, Masatoshi and Ando, Yoichi},
title = {Topological superconductors: a review},
journal = {Rep. Prog. Phys.}
}

@article{alicea2011non,
  title={Non-Abelian statistics and topological quantum information processing in 1D wire networks},
  author={Alicea, Jason and Oreg, Yuval and Refael, Gil and Von Oppen, Felix and Fisher, Matthew P A},
  journal={Nat. Phys.},
  volume={7},
  number={5},
  pages={412--417},
  year={2011},
  doi={https://doi.org/10.1038/nphys1915},
  publisher={Nature Publishing Group UK London}
}

@article{Pb2Pd,
  title = {Type-I superconductivity in single-crystal ${\mathrm{Pb}}_{2}\mathrm{Pd}$},
  author = {Arushi and Motla, K. and Kataria, A. and Sharma, S. and Beare, J. and Pula, M. and Nugent, M. and Luke, G. M. and Singh, R. P.},
  journal = {Phys. Rev. B},
  volume = {103},
  issue = {18},
  pages = {184506},
  numpages = {7},
  year = {2021},
  month = {May},
  publisher = {American Physical Society},
  doi = {10.1103/PhysRevB.103.184506},
  url = {https://link.aps.org/doi/10.1103/PhysRevB.103.184506}
}

@article{fullprof,
title = {Recent advances in magnetic structure determination by neutron powder diffraction},
journal = {Physica B: Condensed Matter},
volume = {192},
number = {1},
pages = {55-69},
year = {1993},
issn = {0921-4526},
doi = {https://doi.org/10.1016/0921-4526(93)90108-I},
author = {Juan Rodríguez-Carvajal}
}

@article{momma2011vesta,
  title={VESTA 3 for three-dimensional visualization of crystal, volumetric and morphology data},
  author={Momma, Koichi and Izumi, Fujio},
  journal={J. Appl. Cryst.},
  volume={44},
  number={6},
  pages={1272--1276},
  year={2011},
  doi={https://doi.org/10.1107/S0021889811038970},
  publisher={International Union of Crystallography}
}

@article{maisuradze2009comparison,
  title={Comparison of different methods for analyzing $\mu$SR line shapes in the vortex state of type-II superconductors},
  author={Maisuradze, A and Khasanov, R and Shengelaya, A and Keller, H},
  journal={J. Phys.: Condens. Matter},
  volume={21},
  number={7},
  pages={075701},
  year={2009},
  doi={10.1088/0953-8984/21/7/075701},
  publisher={IOP Publishing}
}

@article{weber1993flux,
  title = {Magnetic-flux distribution and the magnetic penetration depth in superconducting polycrystalline ${\mathrm{Bi}}_{2}$${\mathrm{Sr}}_{2}$${\mathrm{Ca}}_{1\mathrm{\ensuremath{-}}\mathit{x}}$${\mathrm{Y}}_{\mathit{x}}$${\mathrm{Cu}}_{2}$${\mathrm{O}}_{8+\mathrm{\ensuremath{\delta}}}$ and ${\mathrm{Bi}}_{2\mathrm{\ensuremath{-}}\mathit{x}}$${\mathrm{Pb}}_{\mathit{x}}$${\mathrm{Sr}}_{2}$${\mathrm{CaCu}}_{2}$${\mathrm{O}}_{8+\mathrm{\ensuremath{\delta}}}$},
  author = {Weber, M. and Amato, A. and Gygax, F. N. and Schenck, A. and Maletta, H. and Duginov, V. N. and Grebinnik, V. G. and Lazarev, A. B. and Olshevsky, V. G. and Pomjakushin, V. Yu. and Shilov, S. N. and Zhukov, V. A. and Kirillov, B. F. and Pirogov, A. V. and Ponomarev, A. N. and Storchak, V. G. and Kapusta, S. and Bock, J.},
  journal = {Phys. Rev. B},
  volume = {48},
  issue = {17},
  pages = {13022--13036},
  year = {1993},
  month = {Nov},
  publisher = {American Physical Society},
  doi = {10.1103/PhysRevB.48.13022},
  url = {https://link.aps.org/doi/10.1103/PhysRevB.48.13022}
}

@article{hayano1979kubo,
  title = {Zero-and low-field spin relaxation studied by positive muons},
  author = {Hayano, R. S. and Uemura, Y. J. and Imazato, J. and Nishida, N. and Yamazaki, T. and Kubo, R.},
  journal = {Phys. Rev. B},
  volume = {20},
  issue = {3},
  pages = {850--859},
  numpages = {0},
  year = {1979},
  month = {Aug},
  publisher = {American Physical Society},
  doi = {10.1103/PhysRevB.20.850},
  url = {https://link.aps.org/doi/10.1103/PhysRevB.20.850}
}

@article{Sajilesh2024hfrhge,
author = {Sajilesh, K. P. and Kushwaha, R. K. and Samanta, D. and Tula, T. and Meena, P. K. and Srivastava, S. and Singh, D. and Biswas, P. K. and Kanigel, A. and Hillier, A. D. and Ghosh, S. K. and Singh, R. P.},
title = {Time-Reversal Symmetry Breaking Superconductivity in HfRhGe: A Noncentrosymmetric Weyl Semimetal},
journal = {Adv. Mater.},
volume = {37},
issue = {7},
pages = {2415721},
year = {1979},
doi = {https://doi.org/10.1002/adma.202415721}
}

@article{luke1998time,
  title={Time-reversal symmetry-breaking superconductivity in ${\mathrm{Sr}}_{2}{\mathrm{RuO}}_{4}$},
  author={Luke, G M and Fudamoto, Y and Kojima, K M and Larkin, M I and Merrin, J and Nachumi, B and Uemura, Y J and Maeno, Y and Mao, Z Q and Mori, Y and Nakamura, H. and Sigrist, M.},
  journal={Nature},
  volume={394},
  number={6693},
  pages={558--561},
  year={1998},
  doi={https://doi.org/10.1038/29038},
  publisher={Nature Publishing Group UK London}
}

@article{singh2018re6ti,
  title = {Time-reversal symmetry breaking in the noncentrosymmetric superconductor ${\mathrm{Re}}_{6}\mathrm{Ti}$},
  author = {Singh, D. and Sajilesh, K. P. and Barker, J. A. T. and Paul, D. McK. and Hillier, A. D. and Singh, R. P.},
  journal = {Phys. Rev. B},
  volume = {97},
  issue = {10},
  pages = {100505},
  year = {2018},
  month = {Mar},
  publisher = {American Physical Society},
  doi = {10.1103/PhysRevB.97.100505},
  url = {https://link.aps.org/doi/10.1103/PhysRevB.97.100505}
}

@article{burkov2016topological,
  title={Topological semimetals},
  author={Burkov, A. A.},
  journal={Nat. Mater.},
  volume={15},
  number={11},
  pages={1145--1148},
  year={2016},
  doi={https://doi.org/10.1038/nmat4788},
  publisher={Nature Publishing Group UK London}
}

@article{armitage2018weyl,
  title = {Weyl and Dirac semimetals in three-dimensional solids},
  author = {Armitage, N. P. and Mele, E. J. and Vishwanath, Ashvin},
  journal = {Rev. Mod. Phys.},
  volume = {90},
  issue = {1},
  pages = {015001},
  numpages = {57},
  year = {2018},
  month = {Jan},
  publisher = {American Physical Society},
  doi = {10.1103/RevModPhys.90.015001},
  url = {https://link.aps.org/doi/10.1103/RevModPhys.90.015001}
}

@article{gao2019tsm,
   author = "Gao, Heng and Venderbos, Jörn W. F. and Kim, Youngkuk and Rappe, Andrew M.",
   title = "Topological Semimetals from First Principles", 
   journal= "Annu. Rev. Mater. Res.",
   year = "2019",
   volume = "49",
   number = "Volume 49, 2019",
   pages = "153-183",
   doi = "https://doi.org/10.1146/annurev-matsci-070218-010049",
   url = "https://www.annualreviews.org/content/journals/10.1146/annurev-matsci-070218-010049",
   publisher = "Annual Reviews",
   issn = "1545-4118",
   type = "Journal Article"
}

@article{Ghosh_2021,
doi = {10.1088/1361-648X/abaa06},
url = {https://dx.doi.org/10.1088/1361-648X/abaa06},
year = {2020},
month = {oct},
publisher = {IOP Publishing},
volume = {33},
number = {3},
pages = {033001},
author = {Ghosh, Sudeep Kumar and Smidman, Michael and Shang, Tian and Annett, James F and Hillier, Adrian D and Quintanilla, Jorge and Yuan, Huiqiu},
title = {Recent progress on superconductors with time-reversal symmetry breaking},
journal = {J. Phys.: Condens. Matter}
}

@article{Allen2,
  title = {Transition temperature of strong-coupled superconductors reanalyzed},
  author = {Allen, P. B. and Dynes, R. C.},
  journal = {Phys. Rev. B},
  volume = {12},
  issue = {3},
  pages = {905--922},
  numpages = {0},
  year = {1975},
  month = {Aug},
  publisher = {American Physical Society},
  doi = {10.1103/PhysRevB.12.905},
  url = {https://link.aps.org/doi/10.1103/PhysRevB.12.905}
}

@article{Allen1,
  title = {Neutron Spectroscopy of Superconductors},
  author = {Allen, Philip B.},
  journal = {Phys. Rev. B},
  volume = {6},
  issue = {7},
  pages = {2577--2579},
  numpages = {0},
  year = {1972},
  month = {Oct},
  publisher = {American Physical Society},
  doi = {10.1103/PhysRevB.6.2577},
  url = {https://link.aps.org/doi/10.1103/PhysRevB.6.2577}
}

@article{migdal,
  title={Interaction between electrons and lattice vibrations in a normal metal},
  author={Migdal, A. B.},
  journal={Sov. Phys. JETP},
  volume={7},
  number={6},
  pages={996--1001},
  year={1958},
  doi={http://jetp.ras.ru/cgi-bin/e/index/e/7/6/p996?a=list}
}

@article{DFT1,
  title = {Inhomogeneous Electron Gas},
  author = {Hohenberg, P. and Kohn, W.},
  journal = {Phys. Rev.},
  volume = {136},
  issue = {3B},
  pages = {B864--B871},
  numpages = {0},
  year = {1964},
  month = {Nov},
  publisher = {American Physical Society},
  doi = {10.1103/PhysRev.136.B864},
  url = {https://link.aps.org/doi/10.1103/PhysRev.136.B864}
}

@article{DFT2,
  title = {Self-Consistent Equations Including Exchange and Correlation Effects},
  author = {Kohn, W. and Sham, L. J.},
  journal = {Phys. Rev.},
  volume = {140},
  issue = {4A},
  pages = {A1133--A1138},
  numpages = {0},
  year = {1965},
  month = {Nov},
  publisher = {American Physical Society},
  doi = {10.1103/PhysRev.140.A1133},
  url = {https://link.aps.org/doi/10.1103/PhysRev.140.A1133}
}

@article{PBE,
  title = {Generalized Gradient Approximation Made Simple},
  author = {Perdew, John P. and Burke, Kieron and Ernzerhof, Matthias},
  journal = {Phys. Rev. Lett.},
  volume = {77},
  issue = {18},
  pages = {3865--3868},
  numpages = {0},
  year = {1996},
  month = {Oct},
  publisher = {American Physical Society},
  doi = {10.1103/PhysRevLett.77.3865},
  url = {https://link.aps.org/doi/10.1103/PhysRevLett.77.3865}
}

@article{pseudopotentials1,
  title = {Self-consistent treatment of spin-orbit coupling in solids using relativistic fully separable ab initio pseudopotentials},
  author = {Theurich, Gerhard and Hill, Nicola A.},
  journal = {Phys. Rev. B},
  volume = {64},
  issue = {7},
  pages = {073106},
  numpages = {4},
  year = {2001},
  month = {Jul},
  publisher = {American Physical Society},
  doi = {10.1103/PhysRevB.64.073106},
  url = {https://link.aps.org/doi/10.1103/PhysRevB.64.073106}
}

@article{pseudopotentials2,
  title = {Special points for Brillouin-zone integrations},
  author = {Monkhorst, Hendrik J. and Pack, James D.},
  journal = {Phys. Rev. B},
  volume = {13},
  issue = {12},
  pages = {5188--5192},
  numpages = {0},
  year = {1976},
  month = {Jun},
  publisher = {American Physical Society},
  doi = {10.1103/PhysRevB.13.5188},
  url = {https://link.aps.org/doi/10.1103/PhysRevB.13.5188}
}

@article{DFPT,
  title = {Phonons and related crystal properties from density-functional perturbation theory},
  author = {Baroni, Stefano and de Gironcoli, Stefano and Dal Corso, Andrea and Giannozzi, Paolo},
  journal = {Rev. Mod. Phys.},
  volume = {73},
  issue = {2},
  pages = {515--562},
  numpages = {0},
  year = {2001},
  month = {Jul},
  publisher = {American Physical Society},
  doi = {10.1103/RevModPhys.73.515},
  url = {https://link.aps.org/doi/10.1103/RevModPhys.73.515}
}

@article{QE1,
doi = {10.1088/0953-8984/21/39/395502},
url = {https://doi.org/10.1088/0953-8984/21/39/395502},
year = {2009},
month = {sep},
publisher = {},
volume = {21},
number = {39},
pages = {395502},
author = {Giannozzi, Paolo and Baroni, Stefano and Bonini, Nicola and Calandra, Matteo and Car, Roberto and Cavazzoni, Carlo and Ceresoli, Davide and Chiarotti, Guido L and Cococcioni, Matteo and Dabo, Ismaila and Dal Corso, Andrea and de Gironcoli, Stefano and Fabris, Stefano and Fratesi, Guido and Gebauer, Ralph and Gerstmann, Uwe and Gougoussis, Christos and Kokalj, Anton and Lazzeri, Michele and Martin-Samos, Layla and Marzari, Nicola and Mauri, Francesco and Mazzarello, Riccardo and Paolini, Stefano and Pasquarello, Alfredo and Paulatto, Lorenzo and Sbraccia, Carlo and Scandolo, Sandro and Sclauzero, Gabriele and Seitsonen, Ari P and Smogunov, Alexander and Umari, Paolo and Wentzcovitch, Renata M},
title = {QUANTUM ESPRESSO: a modular and open-source software project for quantum simulations of materials},
journal = {J. Phys.: Condens. Matter}
}

@article{QE2,
doi = {10.1088/1361-648X/aa8f79},
url = {https://doi.org/10.1088/1361-648X/aa8f79},
year = {2017},
month = {oct},
publisher = {IOP Publishing},
volume = {29},
number = {46},
pages = {465901},
author = {Giannozzi, P and Andreussi, O and Brumme, T and Bunau, O and Buongiorno Nardelli, M and Calandra, M and Car, R and Cavazzoni, C and Ceresoli, D and Cococcioni, M and Colonna, N and Carnimeo, I and Dal Corso, A and de Gironcoli, S and Delugas, P and DiStasio, R A and Ferretti, A and Floris, A and Fratesi, G and Fugallo, G and Gebauer, R and Gerstmann, U and Giustino, F and Gorni, T and Jia, J and Kawamura, M and Ko, H-Y and Kokalj, A and Küçükbenli, E and Lazzeri, M and Marsili, M and Marzari, N and Mauri, F and Nguyen, N L and Nguyen, H-V and Otero-de-la-Roza, A and Paulatto, L and Poncé, S and Rocca, D and Sabatini, R and Santra, B and Schlipf, M and Seitsonen, A P and Smogunov, A and Timrov, I and Thonhauser, T and Umari, P and Vast, N and Wu, X and Baroni, S},
title = {Advanced capabilities for materials modelling with Quantum ESPRESSO},
journal = {J. Phys.: Condens. Matter}
}

@article{Eliashberg,
  title={Interactions between electrons and lattice vibrations in a superconductor},
  author={Eliashberg, G M},
  journal={Sov. Phys. JETP},
  volume={11},
  number={3},
  pages={696--702},
  year={1960},
  url={http://www.jetp.ras.ru/cgi-bin/e/index/e/11/3/p696?a=list}
}

@article{ir2ga9,
author = {Wakui ,Kouhei and Akutagawa ,Satoshi and Kase ,Naoki and Kawashima ,Kenji and Muranaka ,Takahiro and Iwahori ,Yasufumi and Abe ,Jiro and Akimitsu ,Jun},
title = {Thermodynamic Properties of the Non-centrosymmetric Type-I Superconductor Rh2Ga9 and Ir2Ga9},
journal = {J. Phys. Soc. Jpn.},
volume = {78},
number = {3},
pages = {034710},
year = {2009},
doi = {10.1143/JPSJ.78.034710},
URL = {https://doi.org/10.1143/JPSJ.78.034710}
}

@book{tinkham2004introduction,
title={Introduction to superconductivity},
author={Tinkham, Michael},
edition={2},
year={1996},
publisher={McGraw–Hill, New York}
}

@article{hillier2022muon,
  title={Muon spin spectroscopy},
  author={Hillier, Adrian D and Blundell, Stephen J and McKenzie, Iain and Umegaki, Izumi and Shu, Lei and Wright, Joseph A and Prokscha, Thomas and Bert, Fabrice and Shimomura, Koichiro and Berlie, Adam and Alberto, Helena and Watanabe, Isao},
  journal={Nat. Rev. Methods Primers},
  volume={2},
  number={1},
  pages={4},
  year={2022},
  doi={https://doi.org/10.1038/s43586-021-00089-0},
  publisher={Nature Publishing Group UK London}
}

@article{mcmillan1968transition,
  title = {Transition Temperature of Strong-Coupled Superconductors},
  author = {McMillan, W. L.},
  journal = {Phys. Rev.},
  volume = {167},
  issue = {2},
  pages = {331--344},
  year = {1968},
  month = {Mar},
  publisher = {American Physical Society},
  doi = {10.1103/PhysRev.167.331},
  url = {https://link.aps.org/doi/10.1103/PhysRev.167.331}
}

@article{padamsee1973quasiparticle,
  title={Quasiparticle phenomenology for thermodynamics of strong-coupling superconductors},
  author={Padamsee, H and Neighbor, J E and Shiffman, C A},
  journal={J. Low Temp. Phys.},
  volume={12},
  pages={387--411},
  year={1973},
  doi = {https://doi.org/10.1007/BF00654872},
  publisher={Springer}
}

@article{Uemura,
  title = {Universal Correlations between ${T}_{c}$ and $\frac{{n}_{s}}{{m}^{*}}$ (Carrier Density over Effective Mass) in High-${T}_{c}$ Cuprate Superconductors},
  author = {Uemura, Y. J. and Luke, G. M. and Sternlieb, B. J. and Brewer, J. H. and Carolan, J. F. and Hardy, W. N. and Kadono, R. and Kempton, J. R. and Kiefl, R. F. and Kreitzman, S. R. and Mulhern, P. and Riseman, T. M. and Williams, D. Ll. and Yang, B. X. and Uchida, S. and Takagi, H. and Gopalakrishnan, J. and Sleight, A. W. and Subramanian, M. A. and Chien, C. L. and Cieplak, M. Z. and Xiao, Gang and Lee, V. Y. and Statt, B. W. and Stronach, C. E. and Kossler, W. J. and Yu, X. H.},
  journal = {Phys. Rev. Lett.},
  volume = {62},
  issue = {19},
  pages = {2317--2320},
  numpages = {0},
  year = {1989},
  month = {May},
  publisher = {American Physical Society},
  doi = {10.1103/PhysRevLett.62.2317},
  url = {https://link.aps.org/doi/10.1103/PhysRevLett.62.2317}
}

@article{hillier1997classification,
  title={The classification of superconductors using muon spin rotation},
  author={Hillier, A D and Cywinski, Robert},
  journal={Appl. Magn. Reson.},
  volume={13},
  pages={95--109},
  year={1997},
  doi={https://doi.org/10.1007/BF03161973},
  publisher={Springer}
}

@misc{SM,
    note = {See Supplementary Material for additional details on methodology, material synthesis, phonon computations, electronic structure calculations, EDX, SAED, magnetization, specific heat, and Uemura plot of SnAs, which includes Refs. \cite{hillier2022muon,DFT1,DFT2,QE1,QE2,PBE,pseudopotentials1,pseudopotentials2,DFPT,Eliashberg,migdal,Allen1,Allen2,ir2ga9,tinkham2004introduction,mcmillan1968transition,padamsee1973quasiparticle,Uemura,hillier1997classification}.}
}

\clearpage

\onecolumngrid

\begin{center}
    {\Large \textbf{Supplementary Material to\\
    "Probing the intermediate state of type-I superconductor SnAs using Muon Spin Spectroscopy"}}
\end{center}

\vspace{10pt}

%%%%%%%%%%%%%%%%%%%%%%%%%%%%%%%%%%%%%%%%%%%%%%%%%%%%%%%%%%%%%%%%%%%%%%%%%%%%%%%%%%%%

\setcounter{figure}{0} % Reset figure counter to start at S1

\renewcommand{\thefigure}{S\arabic{figure}}  % Change figure numbering to S1, S2, etc.

\setcounter{table}{0} % Reset figure counter to start at S1

\renewcommand{\thetable}{S\arabic{table}}  % Change figure numbering to S1, S2, etc.

%%%%%%%%%%%%%%%%%%%%%%%%%%%%%%%%%%%%%%%%%%%%%%%%%%%%%%%%%%%%%%%%%%%%%%%%%%%%%%%%%%%%

In the supporting information, we present additional details of the material synthesis, characterization, data analysis, band structure, and phonon spectrum for SnAs single crystals.

\section*{Methodology}
\subsection*{Experimental Details}
The SnAs single crystals were prepared using the modified Bridgman method with a polycrystalline precursor. The powdered form of the raw elements Sn (4N) and As (4N) was taken in the stoichiometric ratio under a KOREA KIYON argon glove box. The polycrystalline sample was prepared by vacuum sealing the pelletized mixture in a quartz tube. The sealed tube was heated to 823 \si{K} for four days. The polycrystalline sample was then vacuum-sealed in a conical-end quartz tube for single-crystal growth. The tube was heated upright to 1073 \si{K} for two days, followed by slow cooling to 823 \si{K}. The single crystals extracted from the molten chunk were cleavable and shiny, as shown in Fig. \textcolor{blue}{1}(a). The single crystal was powdered to obtain the X-ray diffraction (XRD) spectrum under ambient conditions, using a PANalytical X’Pert diffractometer having a Cu-K$_\alpha$ source. To specify the single-crystalline orientation, we obtained the Laue diffraction pattern using a Photonic-Science Laue diffractometer. The TALOS F200S Transmission Electron Microscope (TEM) operating at 200 \si{kV} was used for Energy Dispersive X-ray (EDX) analysis and Selected Area Electron Diffraction (SAED). The magnetization data were obtained using a Quantum Design (MPMS-3) superconducting quantum interference device magnetometer (SQUID). Heat capacity measurement was performed using the two-tau model with a Quantum Design Physical Property Measurement System (PPMS).

The $\mu$SR experiments were performed with the MUSR spectrometer at the ISIS neutron and muon facility, Rutherford Appleton Laboratory, Oxfordshire, UK. 3 \si{g} of the SnAs single crystals oriented were stacked on a silver holder using diluted GE varnish and mounted in the dilution refrigerator, which operates between 35 \si{mK} and 4.2 \si{K}. $\mu$SR measurements were performed in the zero-field (ZF) and transverse-field (TF) modes. The zero-field was ensured using an active-coil compensation system (error limit 1 \si{\micro T}) to reduce the effects of undesirable fields from the surroundings (see \cite{hillier2022muon} for details on $\mu$SR).

\subsection*{Electronic Band Structure Calculations}
Electronic structure calculations were performed within the framework of density functional theory (DFT) using the Quantum ESPRESSO package \cite{DFT1,DFT2,QE1,QE2}. To treat exchange-correlation interactions among electrons, the generalized gradient approximation (GGA), parametrized by Perdew, Burke, and Ernzerhof (PBE), was used \cite{PBE}. Both scalar and fully relativistic projected augmented pseudo potentials were used to account for spin-orbit coupling (SOC) \cite{pseudopotentials1}. The kinetic energy cutoff point for the plane wave basis was set to 60 \si{Ry}, and the crystal structure was fully relaxed until the forces on all atoms were less than 0.1 \si{meV/}\AA. A Monkhorst k-mesh of $8\times8\times8$ was used to sample the Brillouin zone during relaxation \cite{pseudopotentials2}. Further denser k-meshes of $12\times 12\times 12$ and $24\times 24\times 24$ were used for self-consistent and non-self-consistent calculations, respectively.

To analyze the superconducting properties, the phonon spectrum and phonon DOS were calculated using the linear response scheme implemented in the Quantum ESPRESSO package \cite{DFPT,QE2}. Within this scheme, the second-order derivatives of the total energy are calculated to obtain dynamical matrix elements. Furthermore, the Migdal-Eliashberg approach was employed to assess electron-phonon coupling \cite{Eliashberg}. The integral up to the Fermi surface was performed considering Gaussian smearing with a broadening parameter of 0.05 \si{Ry}. Norm-conserving pseudo-potentials and plane-wave expansion of up to 60 \si{Ry} were used. The dynamical matrices were calculated on the mesh of points q $4\times4\times4$.

\begin{figure*}[t]
\includegraphics[width=0.8\columnwidth]{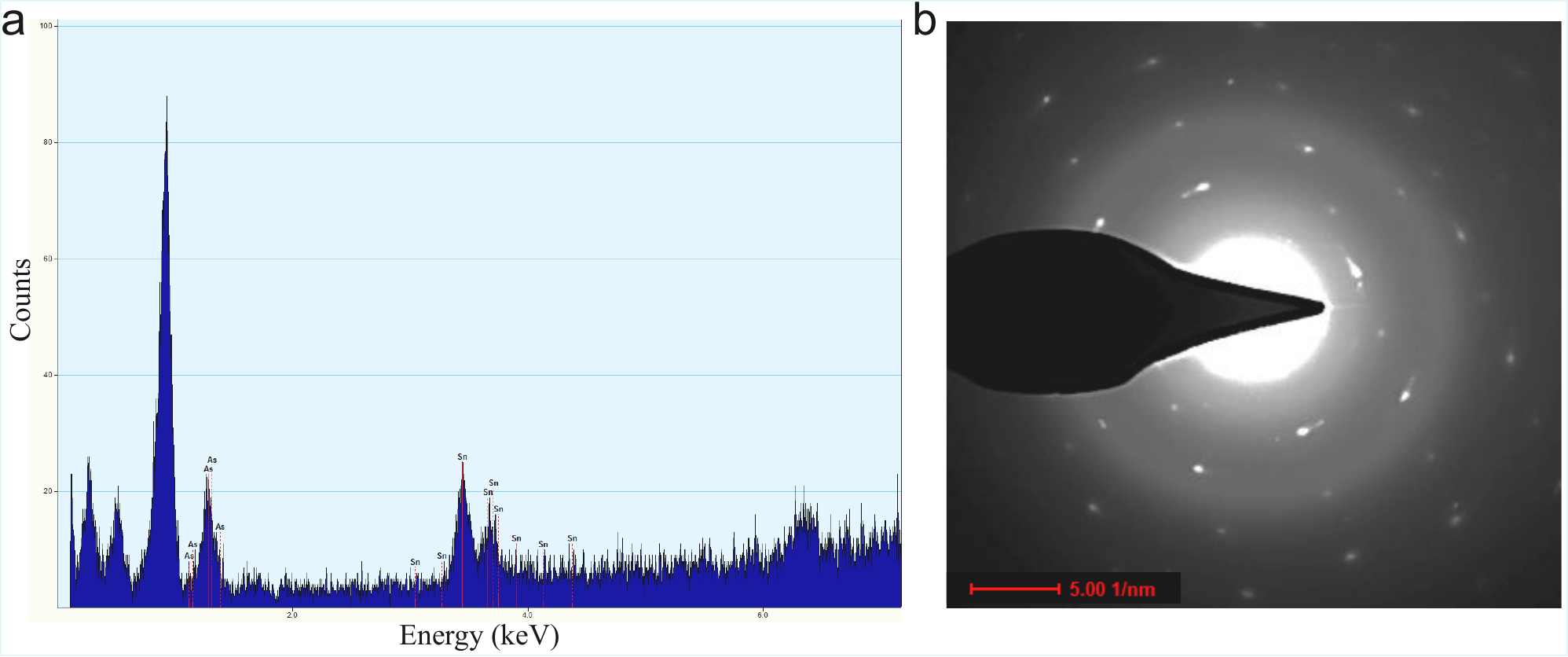}
\caption {\label{Fig:S1}(a) EDX spectra showing Sn and As peaks, and (b) SAED pattern for SnAs single crystal.}
\end{figure*}

\begin{figure*}[b]
\includegraphics[width=\columnwidth]{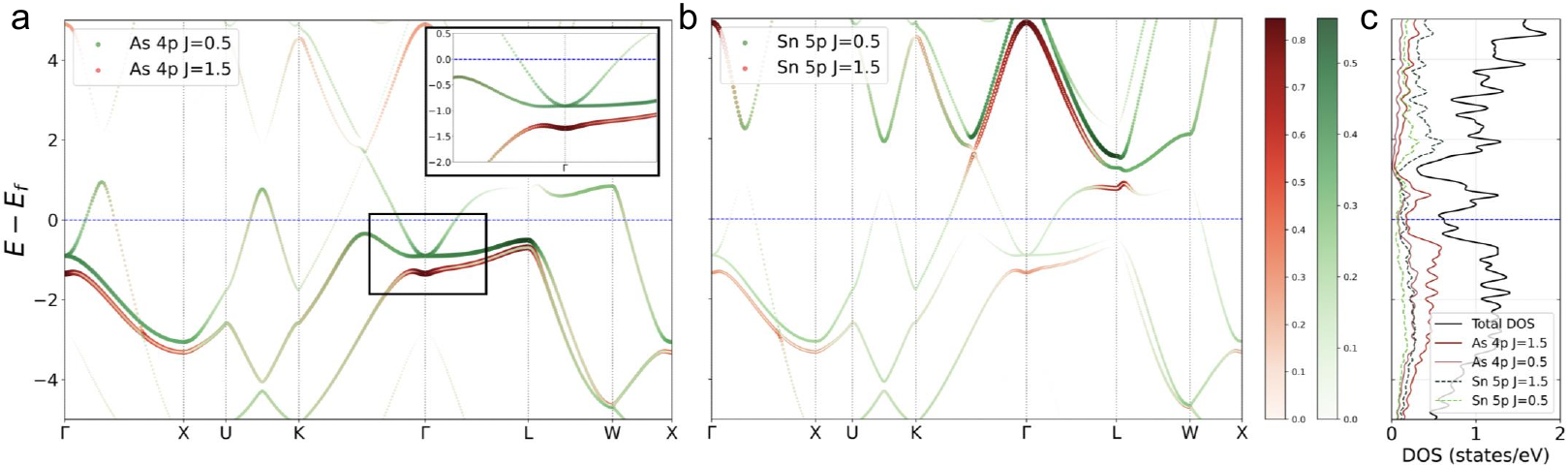}
\caption {\label{Fig:S2} Orbital contribution of (a) As 4p and (b) Sn 5p in the electronic band structure. The inset of (a) reveals the spin contribution in the band degeneracy at the $\Gamma$ point. (c) The orbital- and spin-resolved electronic density of states (DOS).}
\end{figure*}

\section*{\textit{AB INITIO} Computations}
The orbital projections of the As 4p and Sn 5p orbitals over the electronic band structure are shown in Fig. \ref{Fig:S2}(a,b), where the color and thickness of the bands vary according to the orbital weights. The bands at and near $E_{F}$ are predominantly occupied by As 4p, suggesting a strong role for As in electron-phonon coupling. Furthermore, the inset of Fig. \ref{Fig:S2}(a) reveals that the breaking of the degeneracy from six-fold to four-fold at $\Gamma$ is due to spin-orbit coupling. At about 1 \si{eV} above $E_{F}$ in Fig. \ref{Fig:S2}(b), a strong indication of band inversion is seen at the point of high-symmetry L.

\begin{figure}[t]
\includegraphics[width=0.85\columnwidth]{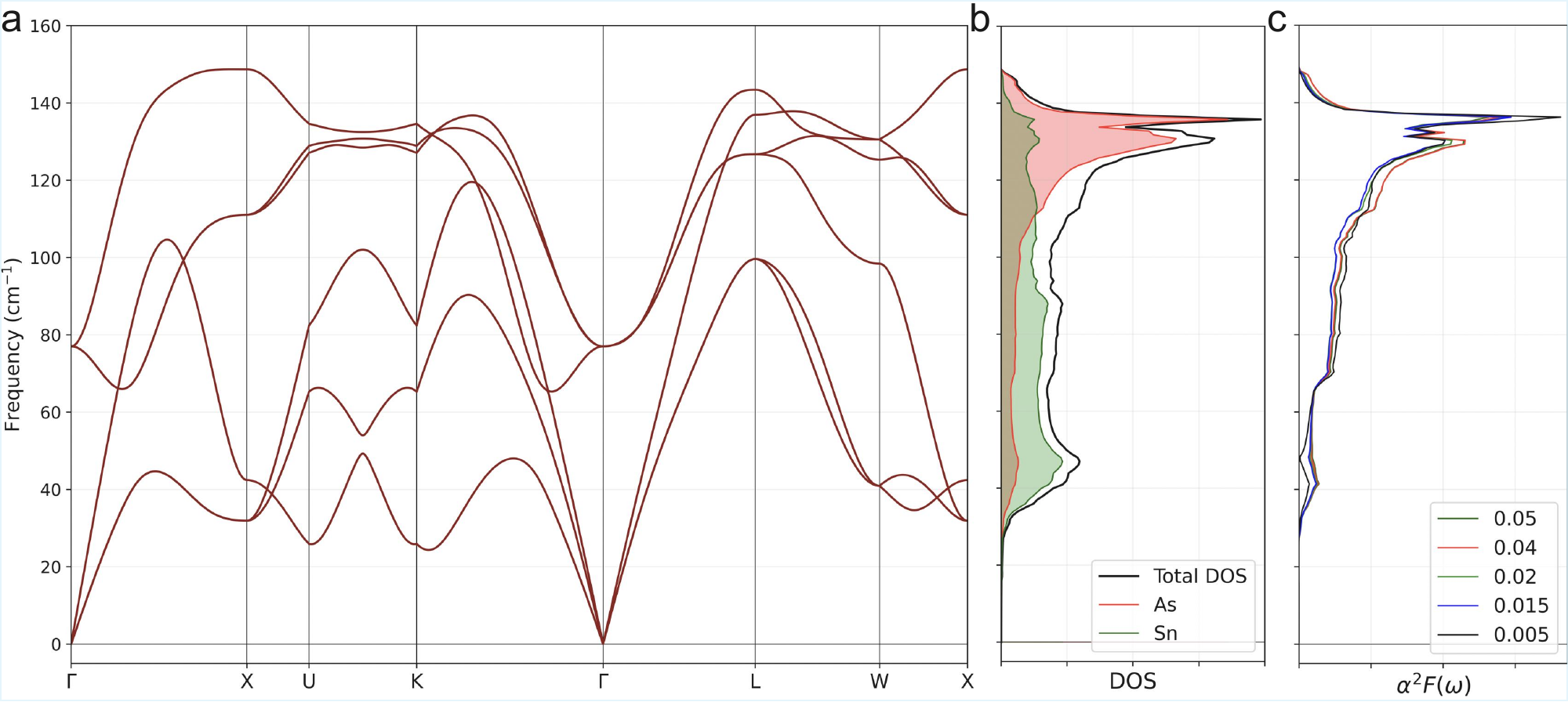}
\caption {\label{Fig:S3}(a) Phonon spectrum and (b) phonon density of states for SnAs. (c) The Eliashberg spectral function remains almost unaffected by varying the numerical broadening.}
\end{figure}

The dispersion of phonons and the density of phonons of states $F(\omega)$ were calculated for SnAs, which is shown in Fig. \ref{Fig:S3}(a,b). The calculations yielded six phonon modes, comprising three acoustic modes and three optical modes. The partial DOS of atomic phonons in Fig. \ref{Fig:S3}(b) shows that As phonons contribute to the higher frequency modes, while lower frequencies are dominated by Sn vibrations. Steep peaks between 130 and 140 \si{cm^{-1}} indicate band flattening in the phonon spectrum.

To verify superconductivity in SnAs, we assumed an electron-phonon pairing mechanism as per the Migdal-Eliashberg theory \cite{migdal,Eliashberg}. The electron-phonon matrix elements were obtained using the Density Functional Perturbation Theory (DFPT) approach implemented in Quantum ESPRESSO. It allowed us to calculate the line widths of the phonons $\gamma_{q\nu}$, where $\nu$ indicates the mode index at a particular q-point. This gives us the local contribution of a phonon to the total electron-phonon coupling parameter $\lambda_{e-ph}$, which in turn can be obtained by summing over the linewidths, $\gamma_{q\nu}$, across all modes and q-points (Eq. \ref{Eq:lambda}).

\begin{figure}[b]
\includegraphics[width=0.7\columnwidth]{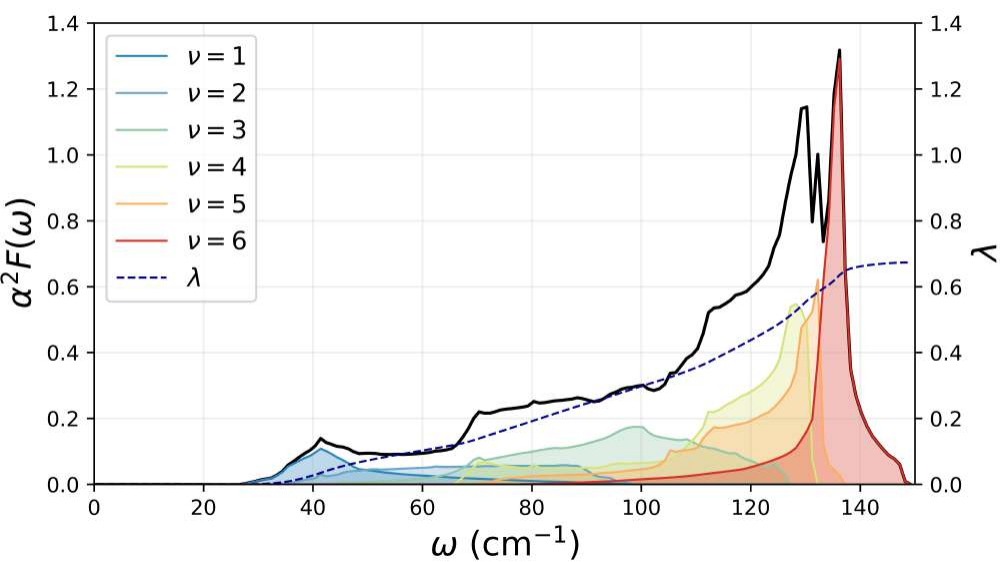}
\caption {\label{Fig:S5}Mode resolved Eliashberg spectral function and average electron-phonon coupling parameter.}
\end{figure}

\begin{equation}
\lambda_{e-ph} = \sum_{q,\nu} \frac{\gamma_{q\nu}}{\pi \hbar (\omega_{q\nu})^{2}N(\epsilon_{F})}
\label{Eq:lambda}
\end{equation}

where $N(\epsilon_{F})$ denotes the total density of states at the Fermi level. This allows for the estimation of the isotropic Eliashberg function by averaging over the Brillouin zone, which gives us an alternative way to evaluate the $\lambda_{e-ph}$ \cite{migdal,Eliashberg,Allen1,Allen2},

\begin{equation}
\alpha^{2}F(\omega) = \frac{1}{2\pi N(\epsilon_{F})} \sum_{q\nu}\frac{\gamma_{q\nu}}{\hbar\omega_{q\nu}}\delta(\omega - \omega_{q\nu}), \quad \lambda_{e-ph} = \int_{0}^{\omega_{max}} \frac{\alpha^{2}F(\omega)}{\omega}d\omega.
\label{Eq:Eliashberg}
\end{equation}

The superconducting transition temperature ($T_{C}$) is calculated using the Allen-Dynes formula \cite{Allen1,Allen2},

\begin{equation}
T_{C} = \frac{\omega_{{\log}}}{1.2} exp\left[\frac{-1.04(1+\lambda)}{\lambda(1-0.62\mu^{*})-\mu^{*}}\right],
\label{Eq:allendynes}
\end{equation}

where the effective screened Coulomb repulsion constant $\mu^{*}$ is taken as 0.13 for intermetallic compounds. $\lambda$ denotes the electron-phonon coupling parameter, and the logarithmic mean frequency, $\omega_{\log}$, is given by Eq. \ref{Eq:frequency}.

\begin{equation}
\omega_{\log} = exp\left[\frac{2}{\lambda}\int\frac{d\omega}{\omega}\alpha^{2}F(\omega)\log\omega\right]
\label{Eq:frequency}
\end{equation}

The Eliashberg spectral functions calculated using different numerical broadenings have an almost similar structure throughout (see Fig. \ref{Fig:S3}(c)). All the computed values of $\lambda$ and $\braket{\omega_{\log}}$ exhibit only a marginal variation, confirming the convergence of our calculations. This leads to the theoretical value of $\lambda_{e-ph}$ being 0.67, which further gives $T_{C}$ 3.56 \si{K}. Following this, Fig. \ref{Fig:S5} shows the corresponding spectral function resolved into its modes, suggesting that the electron-phonon coupling strength is primarily enhanced by the optical branches, corresponding to high-frequency modes. The global integral, which leads to the total $\lambda_{e-ph}$, is also shown in Fig. \ref{Fig:S5}.

\section*{Superconducting properties}

\subsection*{Magnetization}
temperature-dependent magnetization measurements on SnAs single crystals were obtained in two modes: zero-field-cooled warming (ZFCW) and field-cooled cooling (FCC), under a 10 \si{Oe} magnetic field. The superconducting nature of the sample was confirmed from the diamagnetic transition at 3.74(1) \si{K}, which is shown in Fig. \textcolor{blue}{1}(h). The Meissner volume fraction of $>$100\% denotes the complete expulsion of magnetic flux in the superconducting state. This behavior has been observed in a previous report on SnAs \cite{snasprb}, and in many other type-I superconductors, such as \ch{Pb2Pd} \cite{Pb2Pd}, AuBe \cite{aube}, and \ch{Ir2Ge9} \cite{ir2ga9}. Typically, there is a sharp drop in the critical field, which may broaden as a result of the demagnetization effects. The M-T and five-quadrant M-H data give a preliminary indication of the type-I nature of superconductivity in SnAs single crystals. Furthermore, the magnetization versus magnetic field was measured at various temperatures below the superconducting transition temperature ($T_C$) (inset of Fig. \textcolor{blue}{1}(i)). The M-H data at a low magnetic field give the critical magnetic fields at different temperatures, which can be fitted using the Ginzburg-Landau (G-L) equation (Eq. \ref{eq:Hc}) to obtain the value of the critical magnetic field at zero Kelvin, $H_C(0)$ \cite{tinkham2004introduction}.

\begin{equation}
H_{C}(T)=H_{C}(0)[{1-(t)^{2}}],  \quad  \text{where} \;  t = \frac{T}{T_{C}}.
\label{eq:Hc}
\end{equation}

The value of $H_C(0)$ is 187.5(5) \si{Oe}, which can be obtained from the intersection of the G-L fitting curve with the y-axis, as described in Fig. \textcolor{blue}{1}(i).

\begin{table}[b]
\caption{Normal and superconducting state parameters for \ch{SnAs} compound, extracted from magnetization, specific heat, and $\mu$SR measurements.}
\label{tbl:parameters}
\setlength{\tabcolsep}{35pt}
\renewcommand{\arraystretch}{1.3} % Default value: 1
\begin{center}
\resizebox{0.7\columnwidth}{!}{
\begin{tabular}{lcc}\hline \hline
Parameters                                  & unit                  & \ch{SnAs}  \\
\hline
% \\[0.1pt]
$a=b=c$                               & \AA                & 5.7221(5)       \\
$V_{cell}$                               & \AA$^3$                & 187.35(5)       \\
$T_{C}^{mag}$                               & \si{K}                & 3.74(1)       \\
$H_{C}^{mag}(0)$                                 & \si{Oe}               & 187.5(5)       \\
$\gamma_{n}$                                & \si{mJmol^{-1}K^{-2}} & 2.59(1)      \\
$\Theta_{D}$                           & \si{K}                & 204(4)        \\
$\frac{\Delta C_{el}}{\gamma _nT_{C}}$         &  -                    & 1.48(4)       \\
$\frac{\Delta(0)}{k_{B}T_{C}}$         &  -                    & 1.79(5)       \\
$\lambda_{e-ph}$                            & -                     & 0.66(1)       \\
$D_{C}(E_{F})$                              & states/(eV f.u.)      & 1.10(2)       \\
$n$                                         & 10$^{28}$ \si{m^{-3}}       & 6.40(5)       \\
$m^{*}/m_{e}$                               & -                     & 1.25(5)        \\
$T_F$                                       & -                & 53912(29)     \\
$H_{C}^{\mu SR}(0)$                                 & \si{Oe}               & 165(1)       \\
% \\[0.1pt]
\hline \hline
\end{tabular}
}
\par\medskip\footnotesize
\end{center}
\end{table}

\subsection*{Specific Heat}
The bulk superconductivity in SnAs is confirmed using the jump at 3.80(3) \si{K} in the zero-field temperature-dependent specific heat data, as shown in Fig. \textcolor{blue}{1}(j) inset. The low-temperature specific heat data above the superconducting transition were fitted using the Debye-Sommerfeld relation:

\begin{equation}
C=\gamma_{n}T+\beta_{3}T^{3}+\beta_{5}T^{5},
\label{Eq:debye}
\end{equation}

where the first term represents the electronic contribution and the second and third terms indicate the contributions from the lattice and its anharmonicity, respectively. The best fit (inset of Fig. \textcolor{blue}{1}(j)) gives the Sommerfeld coefficient, $\gamma_{n}$ = 2.59(1) \si{mJmol^{-1}K^{-2}}, the phononic constants $\beta_{3}$ = 0.46(1) \si{mJmol^{-1}K^{-4}}, and  $\beta_{5}$ = 0.0005(1) \si{mJmol^{-1}K^{-6}}. The fitting parameters obtained can be used to calculate the Debye temperature $\Theta_{D}$ and the density of states $D_{C}(E_{F})$ at the Fermi energy. The values of $\Theta_{D}$ and $D_{C}(E_{F})$ obtained from Eq. \ref{Eq:theta} are 204(4) \si{K} and 1.10(2) states / (eV f.u.), respectively.

\begin{equation}
\Theta_{D}=\left(\frac{12\pi^{4} R N}{5 \beta_{3}}\right)^{\frac{1}{3}},\quad \gamma_{n} = \left(\frac{\pi^{2} k_{B}^{2}}{3}\right)D_{C}(E_{F}),
\label{Eq:theta}
\end{equation}

where $k_{B}$ and $R$ are the Boltzmann and universal gas constants, and $N$ is the number of atoms in the formula unit of SnAs, respectively. In addition, electron-phonon coupling can be calculated using the inverted McMillan’s equation (Eq. \ref{eqn2:Lambda}) by substituting the value of $\Theta_{D}$ \cite{mcmillan1968transition}. The value of $\lambda_{e-ph}$ = 0.66(1) indicates a weak interaction between electrons and phonons in the superconducting state of SnAs.

\begin{equation}
\lambda_{e-ph} = \frac{1.04+\mu^{*}\mathrm{ln}(\Theta_{D}/1.45T_{C})}{(1-0.62\mu^{*})\mathrm{ln}(\Theta_{D}/1.45T_{C})-1.04}.
\label{eqn2:Lambda}
\end{equation}

The nature of the superconducting gap symmetry can be determined by calculating the temperature-dependent electronic contribution to the specific heat and fitting the data to obtain the best-fitted model, which yields the fitting parameter, mainly the superconducting energy gap value. Subtracting the phononic and anharmonic lattice contributions from the total specific heat yields the electronic specific heat $C_{el}$ (Eq. \ref{Eq:debye}), which is associated with the entropy $S_{el}$ by $C_{el}=t\frac{dS_{el}}{dt}$, where $t=\frac{T}{T_{C}}$. Within the BCS approximation, the superconducting energy gap $\Delta(t)$ can be related to the normalized entropy by

\begin{equation}
\frac{S_{el}}{\gamma_{n} T_{C}}= -\frac{6}{\pi^{2}} \left(\frac{\Delta(0)}{k_{B} T_{C}}\right) \int_{0}^{\infty}[ (1-f).\ln(1-f)+f.\ln(f)]dy,
\label{Eq:swave}
\end{equation}

where, $y=\xi/\Delta(0)$, $\Delta(t) = tanh[1.82\{1.018(\frac{1}{t}-1)\}^{0.51}]$, the Fermi function $f(\xi)=[1+e^{\frac{E(\xi)}{k_{B}T}}]^{-1}$, and the normal electron energy $E(\xi)=\sqrt{\xi^{2}+\Delta^{2}(t)}$. The value of the normalized electronic specific heat jump is $\Delta C_{el}/{\gamma_{n}T_C}$ = 1.48(4), which is close to the limiting value of BCS of 1.43, indicating weakly coupled superconductivity in SnAs. The electronic specific heat data at low temperature fits nicely with the isotropic s-wave model, described in Fig. \textcolor{blue}{1}(j) \cite{padamsee1973quasiparticle}. The superconducting energy gap can be quantified using the fitting, which is given by $\Delta(0)/{k_{B}T_{C}}$ = 1.79(5). The value of $\Delta(0)/{k_{B}T_{C}}$ is comparable to the BCS value of the isotropic energy gap for superconductors, which is 1.76. The values of electron-phonon coupling, specific heat jump, and energy gap suggest weakly coupled superconductivity with an isotropic energy gap.

\subsection*{Uemura Plot}
\begin{figure}[h]
\includegraphics[width=0.6\columnwidth]{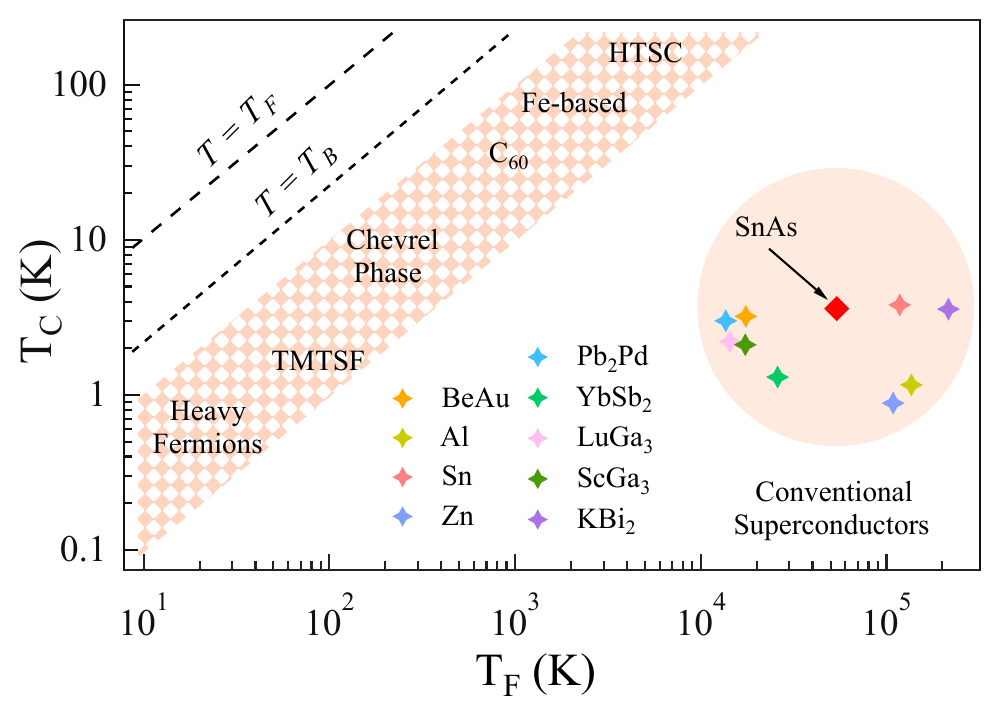}
\caption {\label{Fig:S4} The Uemura plot for classifying the unconventional superconductors. SnAs (red marker) lies in the vicinity of the other type-I superconductors, indicated by the shaded circular region.}
\end{figure}

Uemura modeled a close correlation between the superconducting critical temperature, $T_C$, and the Fermi temperature, $T_F$, in a logarithmic plot, classifying superconductors exhibiting unconventional behavior from those showing conventional behavior. The heavy Fermions, Chevrel phase, Fe-based, and other high $T_C$ unconventional superconductors follow the same linear trend in the ratio between $0.01<T_C/T_F<0.1$ \cite{Uemura}. This region of the Uemera plot is known as the unconventional band shown in Fig. \ref{Fig:S4}.

\begin{equation}
k_{B}T_{F} = \frac{\hbar^{2}}{2m^{*}}{(3\pi^{2}n)^{2/3}}.
\label{eqn:tf}
\end{equation}

The Fermi temperature can be calculated using Eq. \ref{eqn:tf}, where $n$ and $m^*$ represent the carrier density and the effective electron mass, respectively \cite{hillier1997classification}. For the SnAs unit cell that has twelve electrons corresponding to the four formula units, the carrier density is $n=12/V_{cell}=6.40(5)\times10^{28}$ \si{m^{-3}}, which is used to calculate the Fermi wave vector $k_F=(3\pi^2n)^{1/3}=1.23(8)\times10^{-10}$ \si{m^{-1}} \cite{snasvalence}. The effective electron mass $m^{*}=(\hbar k_F)^2\gamma_n/\pi^2nk_B^2$ in terms of electron mass $m_e$ yields the value of $1.25(5)m_{e}$, which further gives the value of $T_F$ as 53912(29) \si{K}. Fig. \ref{Fig:S4} shows some of the type-I superconductors that are found to lie in the circular shaded region, along with SnAs (red marker), far away from the unconventional band.

\end{document}